\documentclass[useAMS,usenatbib]{mn2e}
\usepackage{times}
\usepackage{graphicx}
\usepackage{subfigure}
\usepackage{psfrag}
\usepackage{amsmath}
\usepackage{amsfonts}
\usepackage{alg}
\usepackage{algorithm}
\usepackage{algorithmic}

\def\reff@jnl#1{{\rm#1\/}}
\def\aj{\reff@jnl{AJ}}                 
\def\araa{\reff@jnl{ARA\&A}}           
\def\apj{\reff@jnl{ApJ}}               
\def\apjl{\reff@jnl{ApJ}}              
\def\apjs{\reff@jnl{ApJS}}             
\def\ao{\reff@jnl{Appl.Optics}}        
\def\apss{\reff@jnl{Ap\&SS}}           
\def\aap{\reff@jnl{A\&A}}              
\def\aapr{\reff@jnl{A\&A~Rev.}}        
\def\aaps{\reff@jnl{A\&AS}}            
\def\azh{\reff@jnl{AZh}}               
\def\baas{\reff@jnl{BAAS}}             
\def\jrasc{\reff@jnl{JRASC}}           
\def\memras{\reff@jnl{MmRAS}}          
\def\mnras{\reff@jnl{MNRAS}}           
\def\pra{\reff@jnl{Phys.Rev.A}}        
\def\prb{\reff@jnl{Phys.Rev.B}}        
\def\prc{\reff@jnl{Phys.Rev.C}}        
\def\prd{\reff@jnl{Phys.Rev.D}}        
\def\prl{\reff@jnl{Phys.Rev.Lett}}     
\def\pasp{\reff@jnl{PASP}}             
\def\pasj{\reff@jnl{PASJ}}             
\def\qjras{\reff@jnl{QJRAS}}           
\def\skytel{\reff@jnl{S\&T}}           
\def\solphys{\reff@jnl{Solar~Phys.}}   
\def\sovast{\reff@jnl{Soviet~Ast.}}    
\def\ssr{\reff@jnl{Space~Sci.Rev.}}    
\def\zap{\reff@jnl{ZAp}}               
\def\nat{\reff@jnl{Nature}}            

\title[{\sc MultiNest}: efficient and robust Bayesian inference]
{{\sc MultiNest}: an efficient and
  robust Bayesian inference tool for cosmology and particle physics}
\author[F.~Feroz, M.P.~Hobson \& M.~Bridges] {F.~Feroz\thanks{E-mail:
    f.feroz@mrao.cam.ac.uk}, M.P.~Hobson and M.~Bridges\\ Astrophysics
  Group, Cavendish Laboratory, JJ Thomson Avenue, Cambridge CB3 0HE,
  UK\\}

\date{Accepted ---. Received ---; in original form \today}
\pagerange{\pageref{firstpage}--\pageref{lastpage}}
\pubyear{2008}

\begin{document}
\label{firstpage}
\maketitle

\begin{abstract}
We present further development and the first public release of our multimodal nested sampling algorithm, called
{\sc MultiNest}. This Bayesian inference tool calculates the evidence, with an associated error estimate, and
produces posterior samples from distributions that may contain multiple modes and pronounced (curving)
degeneracies in high dimensions. The developments presented here lead to further substantial improvements in
sampling efficiency and robustness, as compared to the original algorithm presented in Feroz \& Hobson (2008),
which itself significantly outperformed existing MCMC techniques in a wide range of astrophysical inference
problems. The accuracy and economy of the {\sc MultiNest} algorithm is demonstrated by application to two toy
problems and to a cosmological inference problem focussing on the extension of the vanilla $\Lambda$CDM model to
include spatial curvature and a varying equation of state for dark energy.  The {\sc MultiNest} software, which
is fully parallelized using MPI and includes an interface to CosmoMC, is available at {\tt
http://www.mrao.cam.ac.uk/software/multinest/}. It will also be released as part of the SuperBayeS package, for
the analysis of supersymmetric theories of particle physics, at {\tt http://www.superbayes.org}
\end{abstract}


\section{Introduction}\label{sec:intro}

Bayesian analysis methods are already widely used in astrophysics and cosmology, and are now beginning to gain
acceptance in particle physics phenomenology. As a consequence, considerable effort has been made to develop
efficient and robust methods for performing such analyses. Bayesian inference is usually considered to divide
into two categories: parameter estimation and model selection. Parameter estimation is typically performed using
Markov chain Monte Carlo (MCMC) sampling, most often based on the standard Metropolis--Hastings algorithm or its
variants, such as Gibbs' or Hamiltonian sampling (see e.g. \citealt{MacKay}). Such methods can be very
computationally intensive, however, and often experience problems in sampling efficiently from a multimodal
posterior distribution or one with large (curving) degeneracies between parameters, particularly in high
dimensions. Moreover, MCMC methods often require careful tuning of the proposal distribution to sample
efficiently, and testing for convergence can be problematic.  Bayesian model selection has been further hindered
by the even greater computational expense involved in the calculation to sufficient precision of the key
ingredient, the Bayesian evidence (also called the marginalized likelihood or the marginal density of the data). 
As the average likelihood of a model over its prior probability space, the evidence can be used to assign
relative probabilities to different models (for a review of cosmological applications, see
\citealt{Mukherjee06}).  The existing preferred evidence evaluation method, again based on MCMC techniques, is
thermodynamic integration (see e.g.  \citealt{Ruanaidh}), which is extremely computationally intensive but has
been used successfully in astronomical applications (see e.g. \citealt{Hobson03, Marshall03, Slosar03,
Niarchou04, Bassett04, Trotta05, Beltran05, Bridges06a}). Some fast approximate methods have been used for
evidence evaluation, such as treating the posterior as a multivariate Gaussian centred at its peak (see e.g.
\citealt{Hobson02}), but this approximation is clearly a poor one for multimodal posteriors (except perhaps if
one performs a separate Gaussian approximation at each mode). The Savage--Dickey density ratio has also been
proposed (Trotta 2005) as an exact, and potentially faster, means of evaluating evidences, but is restricted to
the special case of nested hypotheses and a separable prior on the model parameters. Various alternative
information criteria for astrophysical model selection are discussed by \citet{Liddle07}, but the evidence
remains the preferred method.

Nested sampling (\citealt{Skilling04}) is a Monte Carlo method targetted at the efficient calculation of the
evidence, but also produces posterior inferences as a by-product. In cosmological applications,
\citet{Mukherjee06} showed that their implementation of the method requires a factor of $\sim 100$ fewer
posterior evaluations than thermodynamic integration. To achieve an improved acceptance ratio and efficiency,
their algorithm uses an elliptical bound containing the current point set at each stage of the process to
restrict the region around the posterior peak from which new samples are drawn. \citet{Shaw07} point out that
this method becomes highly inefficient for multimodal posteriors, and hence introduce the notion of clustered
nested sampling, in which multiple peaks in the posterior are detected and isolated, and separate ellipsoidal
bounds are constructed around each mode.  This approach significantly increases the sampling efficiency. The
overall computational load is reduced still further by the use of an improved error calculation
(\citealt{Skilling04}) on the final evidence result that produces a mean and standard error in one sampling,
eliminating the need for multiple runs. In our previous paper (\citealt{feroz08} -- hereinafter FH08), we built
on the work of \citet{Shaw07} by pursuing further the notion of detecting and characterising multiple modes in
the posterior from the distribution of nested samples, and presented a number of innovations that resulted in a
substantial improvement in sampling efficiency and robustness, leading to an algorithm that constituted a viable,
general replacement for traditional MCMC sampling techniques in astronomical data analysis.

In this paper, we present further substantial development of the method discussed in FH08 and make the first
public release of the resulting Bayesian inference tool, called {\sc MultiNest}. In particular, we propose
fundamental changes to the `simultaneous ellipsoidal sampling' method described in FH08, which result in a
substantially improved and fully parallelized algorithm for calculating the evidence and obtaining posterior
samples from distributions with (an unkown number of) multiple modes and/or pronounced (curving) degeneracies
between parameters. The algorithm also naturally identifies individual modes of a distribution, allowing for the
evaluation of the `local' evidence and parameter constraints associated with each mode separately.

The outline of the paper is as follows. In
Section \ref{sec:bayesian_infer}, we briefly review the basic aspects of
Bayesian inference for parameter estimation and model selection. In
Section \ref{sec:nested}, we introduce nested sampling and discuss the
use of ellipsoidal bounds in Section \ref{sec:ellipsoidal_sampling}. In
Section \ref{sec:improved_ellipsoidal}, we present the {\sc MultiNest}
algorithm. In Section \ref{sec:applications}, we apply our new algorithms
to two toy problems to demonstrate the accuracy and efficiency of the
evidence calculation and parameter estimation as compared with other
techniques. In Section \ref{sec:cosmology}, we consider the use of our
new algorithm for cosmological model selection focussed on the
extension of the vanilla $\Lambda$CDM model to include spatial
curvature and a varying equation of state for dark energy.  We compare
the efficiency of {\sc MultiNest} and standard MCMC techniques for
cosmological parameter estimation in
Section \ref{sec:MCMC_comparison}. Finally, our conclusions are presented
in Section \ref{sec:conclusions}.

\section{Bayesian Inference}\label{sec:bayesian_infer}

Bayesian inference methods provide a consistent approach to the
estimation of a set parameters $\mathbf{\Theta}$ in a model (or
hypothesis) $H$ for the data $\mathbf{D}$. Bayes' theorem states that
\begin{equation} \Pr(\mathbf{\Theta}|\mathbf{D}, H) =
\frac{\Pr(\mathbf{D}|\,\mathbf{\Theta},H)\Pr(\mathbf{\Theta}|H)}
{\Pr(\mathbf{D}|H)},
\end{equation}
where $\Pr(\mathbf{\Theta}|\mathbf{D}, H) \equiv \mathcal{P}(\mathbf{\Theta})$
is the posterior probability distribution of the parameters,
$\Pr(\mathbf{D}|\mathbf{\Theta}, H) \equiv
\mathcal{L}(\mathbf{\Theta})$ is the likelihood,
$\Pr(\mathbf{\Theta}|H) \equiv \pi(\mathbf{\Theta})$ is the prior, and
$\Pr(\mathbf{D}|H) \equiv \mathcal{Z}$ is the Bayesian evidence.

In parameter estimation, the normalising evidence factor is usually
ignored, since it is independent of the parameters $\mathbf{\Theta}$,
and inferences are obtained by taking samples from the (unnormalised)
posterior using standard MCMC sampling methods, where at equilibrium
the chain contains a set of samples from the parameter space
distributed according to the posterior. This posterior constitutes the
complete Bayesian inference of the parameter values, and can be
marginalised over each parameter to obtain individual parameter
constraints.

In contrast to parameter estimation problems, in model selection the
evidence takes the central role and is simply the factor required to
normalize the posterior over $\mathbf{\Theta}$:
\begin{equation}
\mathcal{Z} =
\int{\mathcal{L}(\mathbf{\Theta})\pi(\mathbf{\Theta})}d^D \mathbf{\Theta},
\label{eq:3}
\end{equation} 
where $D$ is the dimensionality of the parameter space.  As the
average of the likelihood over the prior, the evidence automatically
implements Occam's razor: a simpler theory with compact parameter
space will have a larger evidence than a more complicated one, unless
the latter is significantly better at explaining the data.  The
question of model selection between two models $H_0$ and $H_1$ can
then be decided by comparing their respective posterior probabilities
given the observed data set $\mathbf{D}$, as follows
\begin{equation}
\frac{\Pr(H_1|\mathbf{D})}{\Pr(H_0|\mathbf{D})}
=\frac{\Pr(\mathbf{D}|H_1)\Pr(H_1)}{\Pr(\mathbf{D}|
H_0)\Pr(H_0)}
=\frac{\mathcal{Z}_1}{\mathcal{Z}_0}\frac{\Pr(H_1)}{\Pr(H_0)},
\label{eq:3.1}
\end{equation}
where $\Pr(H_1)/\Pr(H_0)$ is the a priori probability ratio for the
two models, which can often be set to unity but occasionally requires
further consideration.

Evaluation of the multidimensional integral \eqref{eq:3} is a
challenging numerical task.  The standard technique of thermodynamic
integration draws MCMC samples not from the posterior directly but
from $\mathcal{L}^{\lambda}\pi$ where $\lambda$ is an inverse
temperature that is slowly raised from $\approx 0$ to $1$ according to
some annealing schedule. It is possible to obtain accuracies of within
0.5 units in log-evidence via this method, but in cosmological model
selection applications it typically requires of order $10^6$ samples
per chain (with around 10 chains required to determine a sampling
error). This makes evidence evaluation at least an order of magnitude
more costly than parameter estimation.

\section{Nested sampling}\label{sec:nested}

Nested sampling (\citealt{Skilling04}) is a Monte Carlo technique aimed
at efficient evaluation of the Bayesian evidence, but also produces
posterior inferences as a by-product. A full discussion of the method
is given in FH08, so we give only a briefly description here,
following the notation of FH08. 

\begin{figure}
\begin{center}
\subfigure[]{\includegraphics[width=0.4\columnwidth]{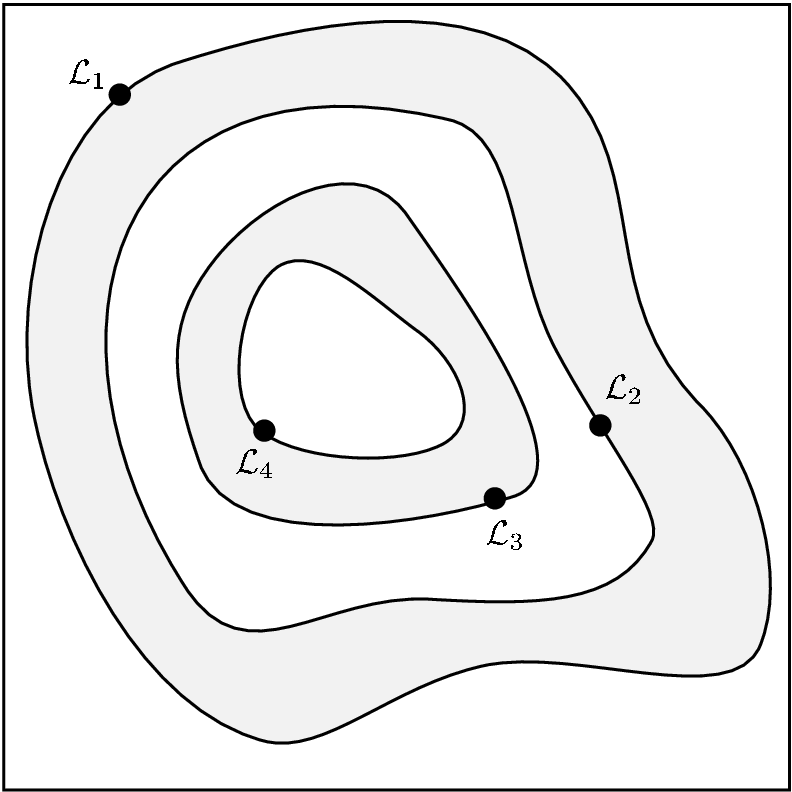}}\hspace{0.3cm}
\subfigure[]{\includegraphics[width=0.4\columnwidth]{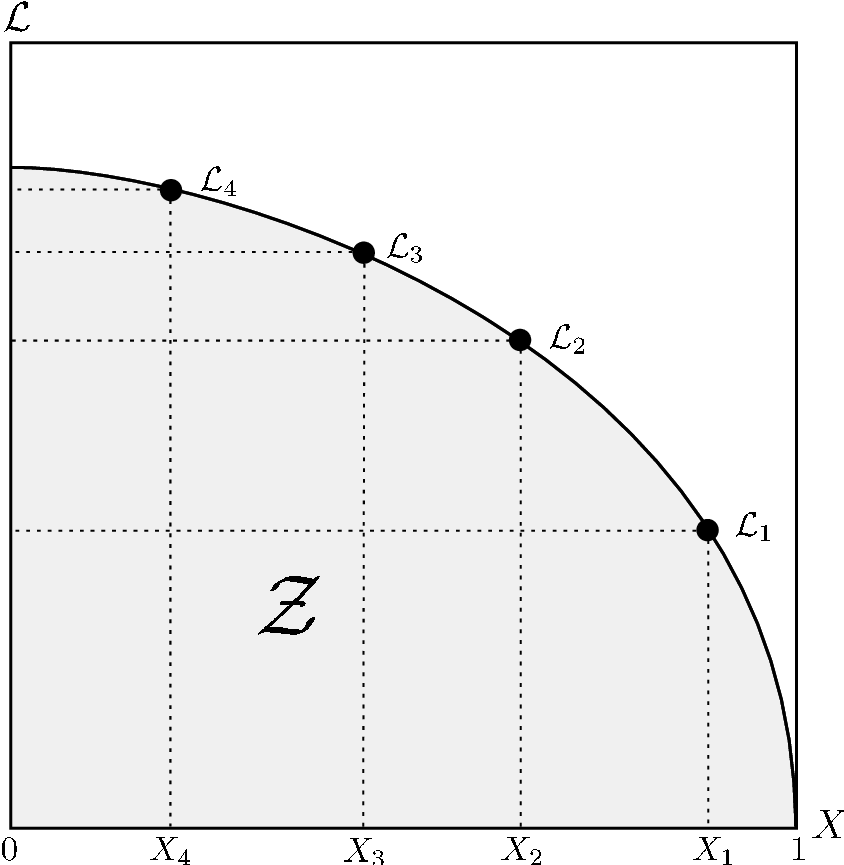}}
\caption{Cartoon illustrating (a) the posterior of a two dimensional problem; and (b) the transformed $\mathcal{L}(X)$ 
function where the prior volumes $X_{i}$ are associated with each likelihood $\mathcal{L}_{i}$.}
\label{figure1}
\end{center}
\end{figure}

Nested sampling exploits the relation
between the likelihood and prior volume to transform the
multidimensional evidence integral (Eq.~\ref{eq:3}) into a one-dimensional
integral. The `prior volume' $X$ is defined by $dX =
\pi(\mathbf{\Theta})d^D \mathbf{\Theta}$, so that
\begin{equation}
X(\lambda) 
= \int_{\mathcal{L}\left(\mathbf{\Theta}\right) > \lambda} 
\pi(\mathbf{\Theta}) d^D\mathbf{\Theta},
\label{Xdef}
\end{equation}
where the integral extends over the region(s) of parameter space
contained within the iso-likelihood contour
$\mathcal{L}(\mathbf{\Theta}) = \lambda$.  
The evidence integral (Eq.~\ref{eq:3}) can then be written as
\begin{equation}
\mathcal{Z}=\int_{0}^{1}{\mathcal{L}(X)}dX,
\label{equation:nested}
\end{equation}
where $\mathcal{L}(X)$, the inverse of Eq.~\ref{Xdef}, is a 
monotonically decreasing function of $X$.  Thus, if one can evaluate
the likelihoods $\mathcal{L}_{i}=\mathcal{L}(X_{i})$, where
$X_{i}$ is a sequence of decreasing values,
\begin{equation}
0<X_{M}<\cdots <X_{2}<X_{1}< X_{0}=1,\label{eq:5}
\end{equation}
as shown schematically in Fig.~\ref{figure1}, the evidence can be
approximated numerically using standard quadrature methods as a
weighted sum
\begin{equation}
\mathcal{Z}={\textstyle {\displaystyle \sum_{i=1}^{M}}\mathcal{L}_{i}w_{i}}.\label{eq:6}
\end{equation} 
In the following we will use the simple trapezium rule, for which the
weights are given by $w_{i}=\frac{1}{2}(X_{i-1}-X_{i+1})$. An
example of a posterior in two dimensions and its associated function
$\mathcal{L}(X)$ is shown in Fig.~\ref{figure1}.

The summation (Eq.~\ref{eq:6}) is performed as follows. The iteration
counter is first set to $i=0$ and $N$ `active' (or `live') samples are
drawn from the full prior $\pi(\mathbf{\Theta})$ (which is often
simply the uniform distribution over the prior range), so the initial
prior volume is $X_{0}=1$.  The samples are then sorted in order of
their likelihood and the smallest (with likelihood $\mathcal{L}_{0}$)
is removed from the active set (hence becoming `inactive')
and replaced by a point drawn from the
prior subject to the constraint that the point has a likelihood
$\mathcal{L}>\mathcal{L}_{0}$. The corresponding prior volume
contained within this iso-likelihood contour will be a random variable
given by $X_{1} = t_{1} X_{0}$, where $t_{1}$ follows the distribution
$\Pr(t) = Nt^{N-1}$ (i.e. the probability distribution for the largest
of $N$ samples drawn uniformly from the interval $[0,1]$). At each
subsequent iteration $i$, the removal of the lowest likelihood
point $\mathcal{L}_{i}$ in the active set, the drawing of a replacement
with $\mathcal{L} > \mathcal{L}_{i}$ and the reduction of the
corresponding prior volume $X_{i}=t_{i} X_{i-1}$ are repeated, until
the entire prior volume has been traversed. The algorithm thus travels
through nested shells of likelihood as the prior volume is reduced.
The mean and standard deviation of $\log t$, which dominates the
geometrical exploration, are $E[\log t]=-1/N$ and $\sigma[\log
  t]=1/N$. Since each value of $\log t$ is independent, after $i$
iterations the prior volume will shrink down such that $\log
X_{i}\approx-(i\pm\sqrt{i})/N$. Thus, one takes $X_{i} =
\exp(-i/N)$. 

The algorithm is terminated on determining the evidence to some
specified precision (we use 0.5 in log-evidence): at iteration $i$,
the largest evidence contribution that can be made by the remaining
portion of the posterior is $\Delta{\mathcal{Z}}_{i} =
\mathcal{L}_{\rm max}X_{i}$, where $\mathcal{L}_{\rm max}$ is the
maximum likelihood in the current set of active points.  The evidence
estimate (Eq.~\ref{eq:6}) may then be refined by adding a final increment
from the set of $N$ active points, which is given by
\begin{equation}
\Delta Z = \sum_{j=1}^N \mathcal{L}_jw_{M+j},
\label{eq:finaldz}
\end{equation}
where $w_{M+j}=X_M/N$ for all $j$.  The final uncertainty on the
calculated evidence may be straightforwardly estimated from a single
run of the nested sampling algorithm by calculating the relative
entropy of the full sequence of samples (see FH08).

Once the evidence $\mathcal{Z}$ is found, posterior inferences can be
easily generated using the full sequence of (inactive and active)
points generated in the nested sampling process.
Each such point is simply assigned the weight
\begin{equation}
p_{j}=\frac{\mathcal{L}_{j} w_{j}}{\mathcal{Z}}.\label{eq:12},
\end{equation}
where the sample index $j$ runs from 1 to $\mathcal{N}=M+N$, the
total number of sampled points. These samples
can then be used to calculate inferences of posterior parameters such
as means, standard deviations, covariances and so on, or to construct
marginalised posterior distributions.

\section{Ellipsoidal nested sampling}\label{sec:ellipsoidal_sampling}

The most challenging task in implementing the nested sampling
algorithm is drawing samples from the prior within the hard constraint
$\mathcal{L}> \mathcal{L}_i$ at each iteration $i$. Employing a naive approach that draws
blindly from the prior would result in a steady decrease in the
acceptance rate of new samples with decreasing prior volume (and
increasing likelihood).

Ellipsoidal nested sampling \citep{Mukherjee06} tries to overcome the
above problem by approximating the iso-likelihood contour $\mathcal{L}=\mathcal{L}_i$ by
a $D$-dimensional ellipsoid determined from the covariance matrix of
the current set of active points. New points are then selected from the
prior within this ellipsoidal bound (usually enlarged slightly by some
user-defined factor) until one is obtained that has a likelihood
exceeding that of the removed lowest-likelihood point. In the limit
that the ellipsoid coincides with the true iso-likelihood contour, the
acceptance rate tends to unity.

\begin{figure*}
\begin{center}
\subfigure[]{\includegraphics[width=0.39\columnwidth]{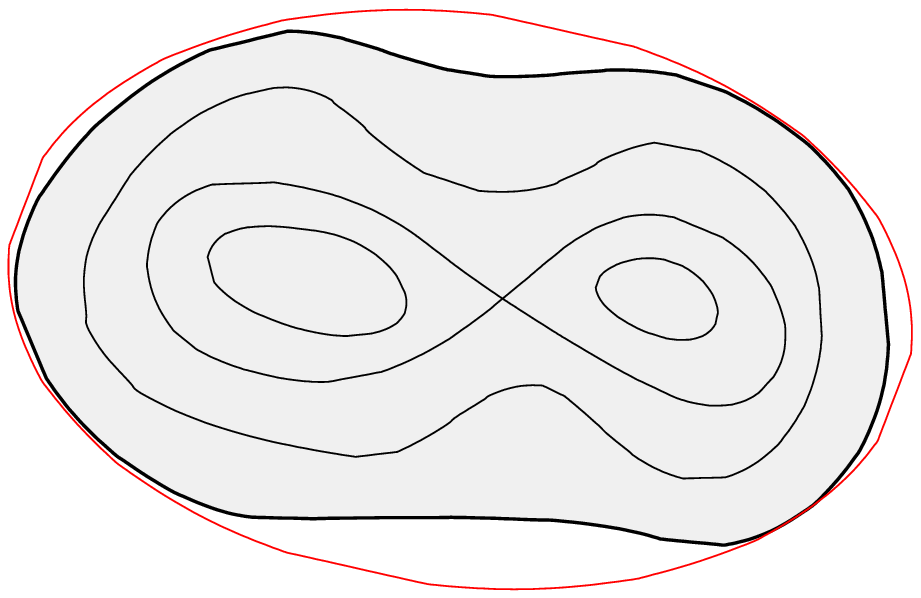}}\setcounter{subfigure}{1}
\subfigure[]{\includegraphics[width=0.39\columnwidth]{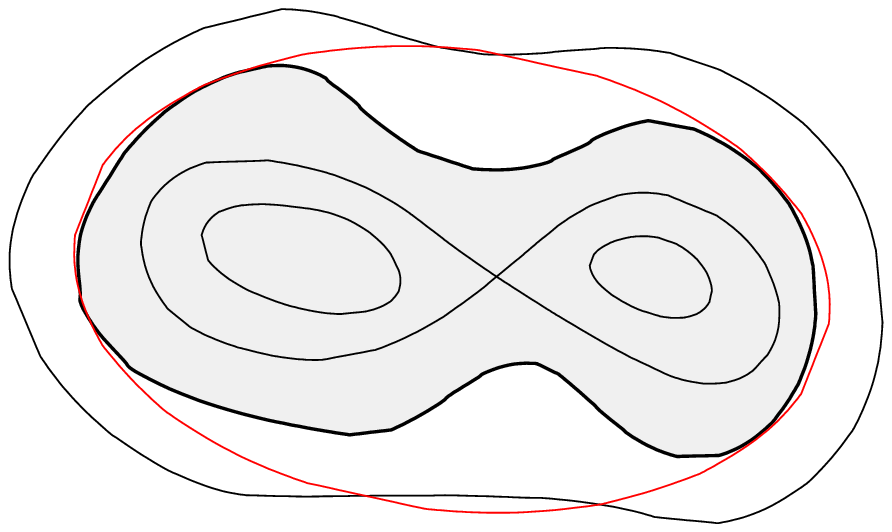}}\setcounter{subfigure}{2}
\subfigure[]{\includegraphics[width=0.39\columnwidth]{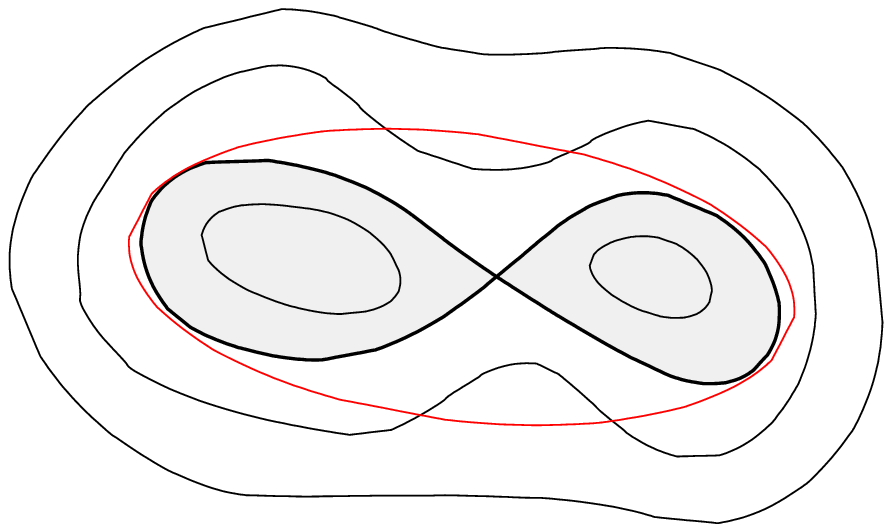}}\setcounter{subfigure}{3}
\subfigure[]{\includegraphics[width=0.39\columnwidth]{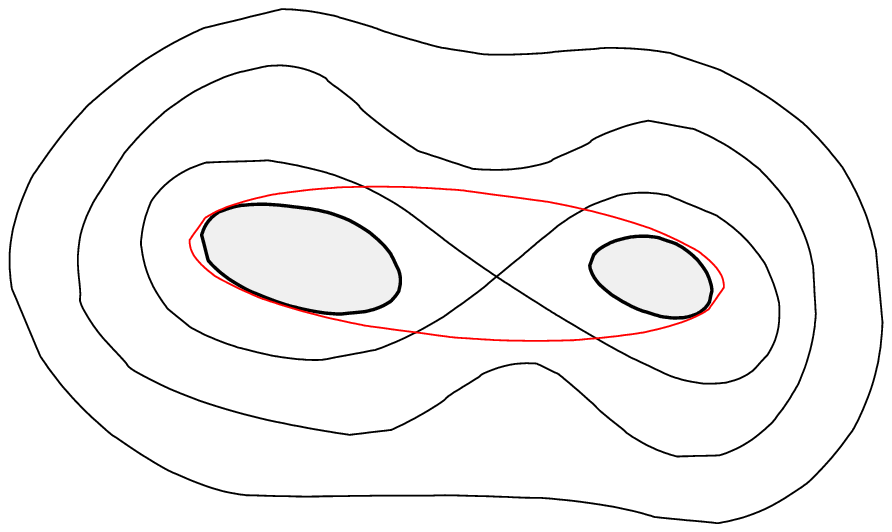}}\setcounter{subfigure}{4} 
\subfigure[]{\includegraphics[width=0.39\columnwidth]{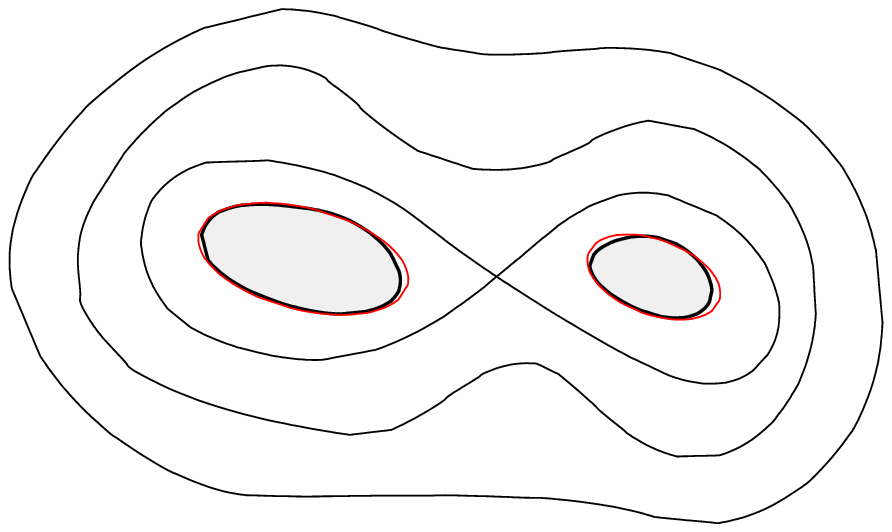}}\setcounter{subfigure}{5} 
\caption{Cartoon of ellipsoidal nested sampling from a simple bimodal
  distribution. In (a) we see that the ellipsoid represents a good
  bound to the active region. In (b)-(d), as we nest inward we can see
  that the acceptance rate will rapidly decrease as the bound steadily
  worsens. Figure (e) illustrates the increase in efficiency obtained
  by sampling from each clustered region separately.}
\label{fig:ellipsoid}
\end{center}
\end{figure*}

Ellipsoidal nested sampling as described above is efficient for simple unimodal posterior distributions without
pronounced degeneracies, but is not well suited to multimodal distributions. As advocated by \citet{Shaw07} and
shown in Fig.~\ref{fig:ellipsoid}, the sampling efficiency can be substantially improved by identifying distinct
\emph{clusters} of active points that are well separated and constructing an individual (enlarged) ellipsoid
bound for each cluster.  In some problems, however, some modes of the posterior may exhibit a pronounced curving
degeneracy so that it more closely resembles a (multi--dimensional) `banana'. Such features are problematic for
all sampling methods, including that of \cite{Shaw07}.

In FH08, we made several improvements to the sampling method of \cite{Shaw07}, which significantly improved its
efficiency and robustness. Among these, we proposed a solution to the above problem by partitioning the set of
active points into as many sub--clusters as possible to allow maximum flexibility in following the degeneracy.
These clusters are then enclosed in ellipsoids and a new point is then drawn from the set of these `overlapping'
ellipsoids, correctly taking into account the overlaps. Although this sub-clustering approach provides maximum
efficiency for highly degenerate distributions, it can result in lower efficiencies for relatively simpler
problems owing to the overlap between the ellipsoids. Also, the factor by which each ellipsoid was enlarged was
chosen arbitrarily. Another problem with the our previous approach was in separating modes with elongated curving
degeneracies. We now propose solutions to all these problems, along with some additional modifications to improve
efficiency and robustness still further, in the {\sc MultiNest} algorithm presented in the following section.

\section{The {\sc MultiNest} algorithm}\label{sec:improved_ellipsoidal}

The {\sc MultiNest} algorithm builds upon the `simultaneous
ellipsoidal nested sampling method' presented in FH08, but
incorporates a number of improvements. In short, at each iteration $i$
of the nested sampling process, the full set of $N$ active points is
partitioned and ellipsoidal bounds constructed using a new algorithm
presented in Section \ref{sec:improved_ellipsoidal:dino} below. This
new algorithm is far more efficient and robust than the method used in
FH08 and automatically accommodates elongated curving degeneracies,
while maintaining high efficiency for simpler problems. This results in
a set of (possibly overlapping) ellipsoids. The lowest-likelihood
point from the full set of $N$ active points is then removed (hence
becoming `inactive') and replaced by a new point drawn from the set of
ellipsoids, correctly taking into account any overlaps. Once a point
becomes inactive it plays no further part in the nested sampling
process, but its details remain stored. We now discuss the {\sc
  MultiNest} algorithm in detail.

\subsection{Unit hypercube sampling space}\label{sec:improved_ellipsoidal:
improvements}

The new algorithm for partitioning the active points into clusters and
constructing ellipsoidal bounds requires the points to be uniformly
distributed in the parameter space.  To satisfy this requirement, the
{\sc MultiNest} `native' space is taken as a $D$-dimensional unit
hypercube (each parameter value varies from 0 to 1) in which samples
are drawn uniformly.  All partitioning of points into clusters,
construction of ellipsoidal bounds and sampling are performed in the
unit hypercube.

In order to conserve probability mass, the point
$\mathbfit{u}=(u_1,u_2,\cdots,u_{D})$ in the unit hypercube should be
transformed to the point
$\mathbf{\Theta}=(\theta_1,\theta_2,\cdots,\theta_{D})$ in the
`physical' parameter space, such that
\begin{equation}
\int \pi(\theta_1, \theta_2, \cdots, \theta_{D})\,d\theta_1
\,d\theta_2 \cdots d\theta_{D} = \int du_1 du_2
\cdots du_{D}.
\label{eq:hypercube1}
\end{equation}
In the simple case that the prior $\pi(\mathbf{\Theta})$ is separable 
\begin{equation}
\pi(\theta_1, \theta_2, \cdots, \theta_{D}) 
= \pi_1(\theta_1)\pi_2(\theta_2)\cdots \pi_{D}(\theta_{D}),
\end{equation}
one can satisfy Eq.~\ref{eq:hypercube1} by setting
\begin{equation}
\pi_{j}(\theta_{j})d\theta_{j} = du_{j}.
\label{eq:hypercube2}
\end{equation}
Therefore, for a given value of $u_{j}$, the corresponding value
of $\theta_{j}$ can be found by solving
\begin{equation}
u_{j} = \int_{-\infty}^{\theta_{j}} \pi_{j}(\theta^\prime_{j}) d\theta^\prime_{j}.
\label{eq:hypercube3}
\end{equation}
In the more general case in which the prior $\pi(\mathbf{\Theta})$ is
not separable, one instead writes
\begin{equation}
\pi(\theta_1, \theta_2, \cdots, \theta_{D}) = \pi_1(\theta_1) \pi_2(\theta_2|\theta_1) \cdots \pi_{D}(\theta_{D}|\theta_1,\theta_2\cdots
\theta_{D-1})
\label{eq:hypercube4}
\end{equation}
where we define
\begin{eqnarray}
\pi_j(\theta_j|\theta_1,\cdots,\theta_{j-1}) & & \nonumber \\
& & \hspace*{-2.7cm} = \int
\pi(\theta_1,\cdots,\theta_{j-1},\theta_j,\theta_{j+1},\cdots,\theta_D)
\,d\theta_{j+1}\cdots d\theta_D.
\label{eq:hypercube6}
\end{eqnarray}
The physical parameters $\mathbf{\Theta}$ corresponding to the
parameters $\mathbfit{u}$ in the unit hypercube can then be found by
replacing the distributions $\pi_j$ in Eq.~\ref{eq:hypercube3} with
those defined in Eq.~\ref{eq:hypercube6} and solving for $\theta_{j}$.
The corresponding physical parameters $\mathbf{\Theta}$ are then used
to calculate the likelihood value of the point
$\mathbfit{u}$ in the unit hypercube. 

It is worth mentioning that in many problems the prior
$\pi(\mathbf{\Theta})$ is uniform, in which case the unit hypercube
and the physical parameter space coincide. Even when this is not so,
one is often able to solve Eq.~\ref{eq:hypercube3} analytically,
resulting in virtually no computational overhead.  For more
complicated problems, two alternative approaches are possible. First,
one may solve Eq.~\ref{eq:hypercube3} numerically, most often using
look-up tables to reduce the computational cost.  Alternatively, one
can re-cast the inference problem, so that the conversion between the
unit hypercube and the physical parameter space becomes trivial. This
is straightforwardly achieved by, for example, defining the new
`likelihood' ${\cal L}^\prime(\mathbf{\Theta})\equiv {\cal L}
(\mathbf{\Theta})\pi(\mathbf{\Theta})$ and `prior'
$\pi^\prime(\mathbf{\Theta}) \equiv \mbox{constant}$. The latter
approach does, however, have the potential to be inefficient since it
does not make use of the true prior $\pi(\mathbf{\Theta})$ to guide
the sampling of the active points.

\subsection{Partitioning and construction of ellipsoidal bounds}\label{sec:improved_ellipsoidal:dino}

In FH08, the partitioning of the set of $N$ active points at each
iteration was performed in two stages. First, X-means (Pelleg et
al. 2000) was used to partition the set into the number of clusters
that optimised the Bayesian Information Criterion (BIC). Second, to
accommodate modes with elongated, curving degeneracies, each cluster
identified by X-means was divided into sub-clusters to follow the
degeneracy. To allow maximum flexibility, this was performed using a
modified, iterative $k$-means algorithm with $k=2$ to produce as many
sub-clusters as possible consistent with there being at least $D+1$
points in any sub-cluster, where $D$ is the dimensionality of the
parameter space. As mentioned above, however, this approach can lead
to inefficiencies for simpler problems in which the iso-likelihood
contour is well described by a few (well-separated) ellipsoidal
bounds, owing to large overlaps between the ellipsoids enclosing each
sub-cluster. Moreover, the factor $f$ by which each ellipsoid was
enlarged was chosen arbitrarily.

We now address these problems by using a new method to partition the
active points into clusters and simultaneously construct the ellipsoidal
bound for each cluster (this also makes redundant the notion of
sub-clustering). At the $i^{\rm th}$ iteration of the nested sampling
process, an `expectation-maximization' (EM) approach is used to find
the optimal ellipsoidal decomposition of $N$ active points distributed
uniformly in a region enclosing prior volume $X_i$, as set out below.

Let us begin by denoting the set of $N$ active points in the unit
hypercube by $S=\{\mathbfit{u}_{1}, \mathbfit{u}_{2}, \cdots,
\mathbfit{u}_{N}\}$ and some partitioning of the set into $K$ clusters
(called the set's $K$-partition) by $\{S_{k}\}^{K}_{k=1}$, where $K
\geq 1$ and $\cup_{k=1}^K S_{k}=S$. For a cluster (or subset)
$S_{k}$ containing $n_{k}$ points, a reasonably
accurate and computationally efficient approximation to its minimum volume
bounding ellipsoid is given by
\begin{equation} 
E_k=\{{\mathbfit{u} \in \mathbb{R}^{D} | \mathbfit{u}^{T} 
(f_k\mathbf{C}_{k})^{-1} \mathbfit{u} \leq 1}\},
\label{eq:ekdef}
\end{equation}
where 
\begin{equation}
\mathbf{C}_{k}=\frac{1}{n_k}\displaystyle\sum_{j=1}^{n_{k}} 
(\mathbfit{u}_{j}-\bmu_{k}) (\mathbfit{u}_{j}-\bmu_{k})^{T}
\label{eq:covariance}
\end{equation}
is the empirical covariance matrix of the subset ${S}_{k}$ and
$\bmu_{k}=\sum_{j=1}^{n_{k}} \mathbfit{u}_{j}$ is its center of the
mass. The enlargement factor $f_k$ ensures that $E_k$ is a bounding
ellipsoid for the subset $S_{k}$. The volume of this ellipsoid,
denoted by $V(E_k)$, is then proportional to
$\sqrt{\det(f_{k}\mathbf{C}_{k})}$. 

Suppose for the moment that we know the volume $V(S)$ of the region
from which the set $S$ is uniformly sampled and let us define the
function
\begin{equation}
F(S) \equiv \frac{1}{V(S)}\sum_{k=1}^K V(E_k).
\label{eq:dino_cost}
\end{equation}
The minimisation of $F(S)$, subject to the constraint $F(S)\ge 1$, with respect to $K$-partitionings
$\{S_k\}_{k=1}^K$ will then yield an `optimal' decomposition into $K$ ellipsoids of the original sampled region. 
The minimisation of $F(S)$ is most easily performed using an `expectation-minimization' scheme as set out below.
This approach makes use of the result \citep{Lu07} that for uniformly distributed points, the variation in $F(S)$
resulting from reassigning a point with position $\mathbfit{u}$ from the subset $S_k$ to the subset $S_{k'}$ is
given by
\begin{equation}
\Delta F(S)_{k,k'} \approx \gamma \left(\frac{V(E_{k'})
d(\mathbfit{u},S_{k'})}{V(S_{k'})} - \frac{V(E_k)
d(\mathbfit{u},S_k)}{V(S_k)} \right)
\label{eq:dino_vary}
\end{equation}
where $\gamma$ is a constant, 
\begin{equation}
d(\mathbfit{u},S_k)=(\mathbfit{u}-\bmu_k)^{T} (f_k
\mathbf{C}_k)^{-1} (\mathbfit{u}-\bmu_k)
\label{eq:mahalanobis}
\end{equation}
is the Mahalanobis distance from $\mathbfit{u}$ to the centroid
$\bmu_k$ of ellipsoid $E_k$ defined in Eq.~\ref{eq:ekdef}, and
\begin{equation}
V(S_k)=\frac{n_kV(S)}{N}
\label{eq:vskdef}
\end{equation}
may be considered as the true volume from which the subset of points
$S_k$ were drawn uniformly.  The approach we have adopted in fact
differs slightly from that outlined above, since we make further use
of Eq.~\ref{eq:vskdef} to impose the constraint that the volume $V(E_k$)
of the $k^{\rm th}$ ellipsoid should never be less than the `true' volume
$V(S_k)$ occupied by the subset $S_k$. This can be easily achieved by
enlarging the ellipsoid $E_k$ by a factor $f_k$, such that its volume
$V(E_k)=\max[V(E_k),V(S_k)]$, before evaluating Eqs.~\ref{eq:dino_cost}
and \ref{eq:dino_vary}.

In our case, however, at each iteration $i$ of the nested sampling
process, $V(S)$ corresponds to the true remaining prior volume $X_i$,
which is not known. Nonetheless, as discussed in
Section \ref{sec:nested}, we do know the expectation value of this
random variable. We thus take $V(S) = \exp(-i/N)$ which, in turn,
allows us to define $V(S_k)$ according to Eq.~\ref{eq:vskdef}. 

From Eq.~\ref{eq:dino_vary}, we see that defining
\begin{equation}
h_k(\mathbfit{u})=\frac{V(E_k) d(\mathbfit{u},S_k)}{V(S_k)}
\label{eq:dino_metric}
\end{equation}
for a point $\mathbfit{u} \in S$ and assigning $\mathbfit{u} \in S_k$
to $S_{k'}$ only if $h_{k}(\mathbfit{u}) < h_{k'}(\mathbfit{u})$,
$\forall$ $k \neq k'$, is equivalent to minimizing $F(S)$ using the
variational formula (Eq.~\ref{eq:dino_vary}). Thus, a weighted Mahalanobis
metric can be used in the $k$-means framework to optimize the
functional $F(S)$. In order to find out the optimal number of
ellipsoids, $K$, a recursive scheme can be used which starts with
$K=2$, optimizes this $2$-partition using the metric in
Eq.~\ref{eq:dino_metric} and recursively partitions the resulting
ellipsoids. For each iteration of this recursion, we employ this
optimization scheme in Algorithm~\ref{alg:dino}.

\begin{algorithm}
  \caption{Minimizing $F(S)$, subject to $F(S) \ge 1$, for points
    $S=\{\mathbfit{u}_1,\mathbfit{u}_2,\cdots,\mathbfit{u}_N\}$
    uniformly distributed in a volume $V(S)$.}
  \label{alg:dino}
  \begin{algorithmic}[1]

    \STATE calculate bounding ellipsoid $E$ and its volume $V(E)$
    \STATE enlarge $E$ so that $V(E)= \max[V(E),V(S)]$. 
    \STATE partition $S$ into $S_1$ and $S_2$ containing $n_1$
    and $n_2$ points
      respectively by applying $k-$means clustering algorithm with $K=2$. 
    \STATE calculate $E_1$, $E_2$ and their volumes $V(E_1)$ and $V(E_2)$
    respectively.
    
    \STATE  enlarge $E_k$ $(k=1,2)$ so that $V(E_k)=\max[V(E_k),V(S_k)]$.
    \FORALL {$\mathbfit{u} \in S$}
    	\STATE assign $\mathbfit{u}$ to $S_k$ such that
        $h_k(\mathbfit{u}) = \min[h_{1}(x), h_{2}(x)]$.
    \ENDFOR
    \IF {no point has been reassigned}
    	\STATE go to step 14.
    \ELSE
    	\STATE go to step 4.
    \ENDIF 
    \IF {$V(E_1) + V(E_2) < V(E)$ or $V(E)>2V(S)$}
      \STATE partition $S$ into $S_{1}$ and $S_2$ and repeat
      entire algorithm for each subset $S_1$ and $S_2$.  
    \ELSE
    	\RETURN $E$ as the optimal ellipsoid of the point set $S$.
    \ENDIF
  \end{algorithmic} 
\end{algorithm}
\begin{figure}
\begin{center}
\subfigure[]{\includegraphics[width=1.1\columnwidth,height=0.8\columnwidth]{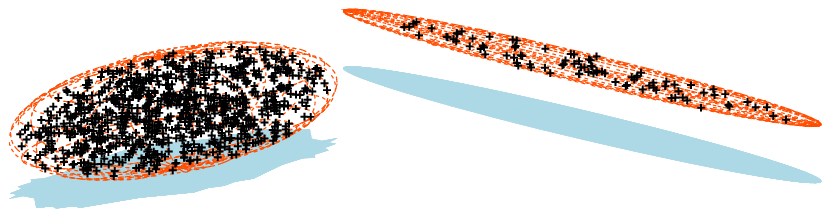}}\setcounter{subfigure}{1}
\subfigure[]{\includegraphics[width=1.1\columnwidth,height=0.8\columnwidth]{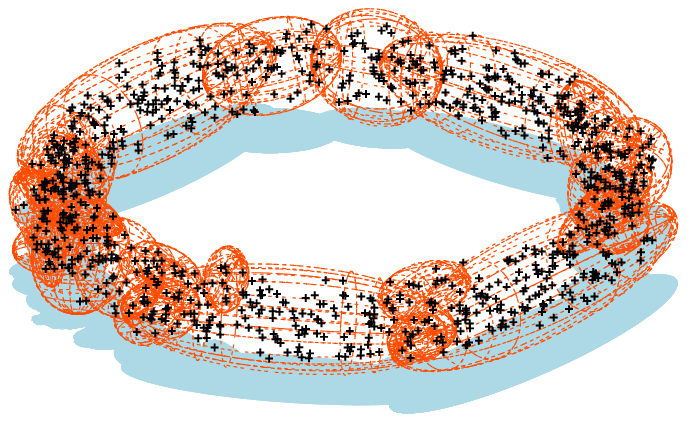}}\setcounter{subfigure}{2}
\caption{Illustrations of the ellipsoidal decompositions returned by
  Algorithm~\ref{alg:dino}: the points given as input are overlaid on
  the resulting ellipsoids. 1000 points
  were sampled uniformly from: (a) two non-intersecting
  ellipsoids; and (b) a torus.}
\label{fig:dino}
\end{center}
\end{figure}

In step 14 of Algorithm~\ref{alg:dino} we partition the point set $S$ with ellipsoidal volume $V(E)$ into subsets
$S_1$ and $S_2$ with ellipsoidal volumes $V(E_1)$ and $V(E_2)$ even if $V(E_1)+V(E_2) > V(E)$, provided $V(E) >
2V(S)$. This is required since, as discussed in \citet{Lu07}, the minimizer of $F(S)$ can be over-conservative
and the partition should still be performed if the ellipsoidal volume is greater than the true volume by some
factor (we use 2).

The above EM algorithm can be quite computationally expensive, especially in higher dimensions, due to the number
of eigenvalue and eigenvector evaluations required to calculate ellipsoidal volumes. Fortunately, {\sc MultiNest}
does not need to perform the full partitioning algorithm at each iteration of the nested sampling process. Once
partitioning of the active points and construction of the ellipsoidal bounds has been performed using
Algorithm~\ref{alg:dino}, the resulting ellipsoids can then be evolved through scaling at subsequent iterations
so that their volumes are $\max[V(E_k),X_{i+1}n_k/N]$, where with $X_{i+1}$ is the remaining prior volume in the
next nested sampling iteration and $n_k$ is number of points in the subset $S_k$ at the end of $i^{\rm th}$
iteration. As the {\sc MultiNest} algorithm moves to higher likelihood regions, one would expect the ellipsoidal
decomposition calculated at some earlier iteration to become less optimal. We therefore perform a full
re-partitioning of the active points using Algorithm~\ref{alg:dino} if $F(S) \geq h$; we typically use $h=1.1$.

The approach outlined above allows maximum flexibility and sampling efficiency by breaking up a posterior mode
resembling a Gaussian into relatively few ellipsoids, but a mode possesses a pronounced curving degeneracy into a
relatively large number of small `overlapping' ellipsoids. In Fig.~\ref{fig:dino} we show the results of applying
Algorithm~\ref{alg:dino} to two different problems in three dimensions: in (a) the iso-likelihood surface
consists of two non-overlapping ellipsoids, one of which contains correlations between the parameters; and in (b)
the iso-likelihood surface is a torus. In each case, 1000 points were uniformly generated inside the
iso-likelihood surface are used as the starting set $S$ in Algorithm~\ref{alg:dino}. In case (a),
Algorithm~\ref{alg:dino} correctly partitions the point set in two non-overlapping ellipsoids with $F(S)=1.1$,
while in case (b) the point set is partitioned into 23 overlapping ellipsoids with $F(S)=1.2$.

In our nested sampling application, it is possible that the ellipsoids found by Algorithm~\ref{alg:dino} might
not enclose the entire iso-likelihood contour, even though the sum of their volumes is constrained to exceed the
prior volume $X$ This is because the ellipsoidal approximation to a region in the prior space might not be
perfect.  It might therefore be desirable to sample from a region with volume greater than the prior volume. This
can easily be achieved by using $X/e$ as the desired minimum volume in Algorithm~\ref{alg:dino}, where $X$ is the
prior volume and $e$ the desired sampling efficiency ($1/e$ is the enlargement factor). We also note that if the
desire sampling efficiency $e$ is set to be greater than unity, then the prior can be under-sampled. Indeed,
setting $e>1$ can be useful if one is not interested in the evidence values, but wants only to have a general
idea of the posterior structure in relatively few likelihood evaluations. We note that, regardless of the value
of $e$, it is always ensured that the ellipsoids $E_k$ enclosing the subsets $S_k$ are always the bounding
ellipsoids.

\subsection{Sampling from overlapping ellipsoids}\label{sec:improved_ellipsoidal:sample_overlap}

Once the ellipsoidal bounds have been constructed at some iteration of
the nested sampling process, one must then draw a new point uniformly
from the union of these ellipsoids, many of which may be overlapping.
This is achieved using the method presented in FH08, which is
summarised below for completeness.

Suppose at iteration $i$ of the nested sampling algorithm, one has
$K$ ellipsoids $\{E_k\}$. One ellipsoid is then chosen with
probability $p_k$ equal to its volume fraction
\begin{equation} p_k=V(E_k)/V_{\rm tot}, 
\label{eq:sample_overlap} 
\end{equation} 
where $V_{\rm tot} = \sum_{k=1}^{K} V(E_k)$. Samples are then drawn
  uniformly from the chosen ellipsoid until a sample is found for
  which the hard constraint $\mathcal{L}>\mathcal{L}_{i}$ is
  satisfied, where $\mathcal{L}_{i}$ is the lowest-likelihood value
  among all the active points at that iteration.  There is, of course, a
  possibility that the chosen ellipsoid overlaps with one or more
  other ellipsoids. In order to take an account of this possibility,
  we find the number of ellipsoids, $n_{e}$, in which the sample
  lies and only accept the sample with probability $1/n_{e}$. This
  provides a consistent sampling procedure in all cases.

\subsection{Decreasing the number of active points}\label{sec:improved_ellipsoidal:dactive}

For highly multimodal problems, the nested sampling algorithm would
require a large number $N$ of active points to ensure that all the modes
are detected. This would consequently result in very slow convergence
of the algorithm. In such cases, it would be desirable to decrease the
number of active points as the algorithm proceeds to higher likelihood
levels, since the number of isolated regions in the iso-likelihood
surface is expected to decrease with increasing likelihood.  modes
Fortunately, nested sampling does not require the number of active
points to remain constant, provided the fraction by which the prior
volume is decreased after each iteration is adjusted accordingly. Without
knowing anything about the posterior, we can use the largest evidence
contribution that can be made by the remaining portion of the
posterior at the $i^{\rm th}$ iteration $\Delta{\mathcal{Z}}_i
= \mathcal{L}_{\rm max}X_{i}$, as the guide in reducing the number
of active points by assuming that the change in $\Delta{\mathcal{Z}}$ is
linear locally. We thus set the number of active points $N_{i}$ at
the $i^{\rm th}$ iteration to be
\begin{equation}
N_{i} = N_{i-1}-N_{\rm min} \frac{\Delta{\mathcal{Z}}_{i-1}
-\Delta{\mathcal{Z}}_{i}}{\Delta{\mathcal{Z}}_{i}-\mbox{tol}}, 
\end{equation}
subject to the constraint $N_{\rm min} \leq N_{i} \leq N_{i-1}$,
where $N_{\rm min}$ is the minimum number of active points allowed and
$\mbox{tol}$ is the tolerance on the final evidence used in the
stopping criterion.

\subsection{Parallelization}\label{sec:improved_ellipsoidal:parallelization}

Even with the enlargement factor $e$ set to unity (see
Section \ref{sec:improved_ellipsoidal:dino}), the typical sampling
efficiency obtained for most problems in astrophysics and particle
physics is around 20--30 per cent for two main reasons. First, the
ellipsoidal approximation to the iso-likelihood surface at any
iteration is not perfect and there may be regions of the parameter
space lying inside the union of the ellipsoids but outside the true
iso-likelihood surface; samples falling in such regions will be
rejected, resulting in a sampling efficiency less than unity. Second,
if the number of ellipsoids at any given iteration is greater than
one, then they may overlap, resulting in some samples with
$\mathcal{L}>\mathcal{L}_{i}$ falling inside a region shared by
$n_{e}$ ellipsoids; such points are accepted only with probability
$1/n_{e}$, which consequently lowers the sampling efficiency.
Since the sampling efficiency is typically less than unity, the {\sc
  MultiNest} algorithm can be usefully (and easily) parallelized by,
at each nested sampling iteration, drawing a potential replacement
point on each of $N_{\rm CPU}$ processors, where $1/N_{\rm CPU}$ is an
estimate of the sampling efficiency.

\subsection{Identification of modes}\label{sec:improved_ellipsoidal:mode_identification}

%
%

As discussed in FH08, for multimodal posteriors it can prove useful to
identify which samples `belong' to which mode. There is inevitably
some arbitrariness in this process, since modes of the posterior
necessarily sit on top of some general `background' in the probability
distribution. Moreover, modes lying close to one another in the
parameter space may only `separate out' at relatively high likelihood
levels. Nonetheless, for well-defined, `isolated' modes, a reasonable
estimate of the posterior mass that each contains (and hence the
associated `local' evidence) can be defined, together with the
posterior parameter constraints associated with each mode.  To perform
such a calculation, once the nested sampling algorithm has progressed
to a likelihood level such that (at least locally) the `footprint' of
the mode is well-defined, one needs to identify at each subsequent
iteration those points in the active set belonging to that mode.  The
partitioning and ellipsoids construction algorithm described in
Section \ref{sec:improved_ellipsoidal:dino} provides a much more
efficient and reliable method for performing this identification, as
compared with the methods employed in FH08.

At the beginning of the nested sampling process, all the active points
are assigned to one `group' $G_1$. As outlined above, at subsequent
iterations, the set of $N$ active points is partitioned into $K$
subsets $\{S_k\}$ and their corresponding ellipsoids $\{E_k\}$
constructed.  To perform mode identification, at each iteration, one
of the subsets $S_k$ is then picked at random: its members become the
first members of the `temporary set' $\mathcal{T}$ and its associated
ellipsoid $E_k$ becomes the first member of the set of ellipsoids
$\mathcal{E}$.  All the other ellipsoids $E_{k'}$ ($k' \neq k)$ are
then checked for intersection with $E_k$ using an exact algorithm
proposed by Alfano \& Greer (2003). Any ellipsoid found to intersect
with $E_k$ is itself added to ${\cal E}$ and the members of the
corresponding subset $S_k$ are added to $\mathcal{T}$.  The set ${\cal
  E}$ (and consequently the set $\mathcal{T}$) is then iteratively
built up by adding to it any ellipsoid not in ${\cal E}$ that
intersects with any of those already in ${\cal E}$, until no more
ellipsoids can be added. Once this is completed, if no more ellipsoids
remain then all the points in $\mathcal{T}$ are (re)assigned to $G_1$,
a new active point is drawn from the union of the ellipsoids $\{E_k\}$
(and also assigned to $G_1$) and the nested sampling process proceeds
to its next iteration.

If, however, there remain ellipsoids $\{E_k\}$ not belonging to ${\cal
  E}$, then this indicates the presence of (at least) two isolated
regions contained within the iso-likelihood surface. In this event,
the points in $\mathcal{T}$ are (re)assigned to the group $G_2$ and
the remaining active points are (re)assigned to the group $G_3$. The
original group $G_1$ then contains only the inactive points generated
up to this nested sampling iteration and is not modified further.  The
group $G_3$ is then analysed in a similar manner to see if can be
split further (into $G_3$ and $G_4$), and the process continued until
no further splitting is possible.  Thus, in this case, one is left
with an `inactive' group $G_1$ and a collection of `active' groups
$G_2$, $G_3$, \ldots. A new active point is then drawn from the union
of the ellipsoids $\{E_k\}$, and assigned to the appropriate active
group, and the nested sampling process proceeds to its next iteration.

At subsequent nested sampling iterations, each of the active groups
present at that iteration is analysed in the same manner to see if it
can be split. If so, then the active points in that group are
(re)assigned to new active groups generated as above, and the original
group becomes inactive, retaining only those of its points that are
inactive.  In order to minimize the computational cost, we take
advantage of the fact that the ellipsoids created by
Algorithm~\ref{alg:dino} can only get smaller in later
iterations. Hence, within each active group, if two ellipsoids are
found not to intersect at some iteration, they are not checked for
intersection in later iterations.  This makes the computational
expense involved in separating out the modes negligible.

At the end of the nested sampling process, one thus obtains a set of
inactive groups and a set of active groups, which between them
partition the full set of (inactive and active) sample points
generated.  It is worth noting that, as the nested sampling process
reaches higher likelihood levels, the number of active points in any
particular active group may dwindle to zero, but such a group is still
considered active since it remains unsplit at the end of the nested
sampling run.  Each active groups is then promoted to a `mode',
resulting in a set of $L$ (say) such modes $\{M_l\}$.

\begin{figure}
\begin{center}
\includegraphics[width=0.7\columnwidth]{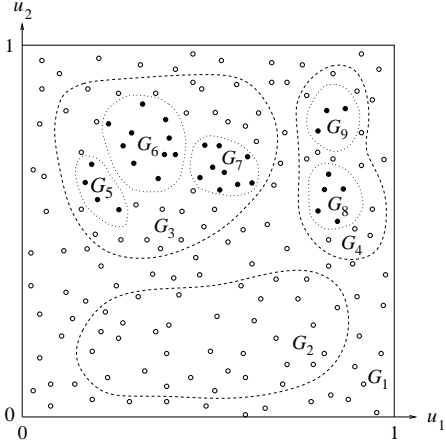}
\caption{Cartoon illustrating the assignment of points to groups; see text
  for details. The iso-likelihood contours
$\mathcal{L}=\mathcal{L}_{i_1}$ and $\mathcal{L}=\mathcal{L}_{i_2}$ are shown as the dashed lines and 
dotted lines respectively. The solid circles denote active points at
the nested sampling iteration $i=i_2$, and the open circles
are the inactive points at this stage.}
\label{fig:groups}
\end{center}
\end{figure}

As a concrete example, consider the two-dimensional illustration shown
in Fig.~\ref{fig:groups}, in which the solid circles denote active
points at the nested sampling iteration $i=i_2$, and the open circles
are the inactive points at this stage.
In this illustration, the first group $G_1$ remains unsplit until
iteration $i=i_1$ of the nested sampling process, at which stage it is
split into $G_2$, $G_3$ and $G_4$. The group $G_3$ then remains unsplit until
iteration $i=i_2$, when it is split into $G_5$, $G_6$ and $G_7$.  
The group $G_4$ remains unsplit until iteration $i=i_2$, when it is
split into $G_8$ and $G_9$. The
group $G_2$ remains unsplit at iteration $i=i_2$ but the number of
active points it contains has fallen to zero, since it is a low-lying
region of the likelihood function. Thus, at the iteration $i=i_2$, the
inactive groups are $G_1$, $G_3$ and $G_4$, and the active groups are $G_2,
G_5, G_6, G_7, G_8$ and $G_6$. If (say) all of the latter collection of groups
were to remain active until the end of the nested sampling process,
each would then be promoted to a mode according to $M_1=G_2$,
$M_2=G_5$, $M_3=G_6,\cdots, M_6=G_9$.

\subsection{Evaluating `local' evidences}\label{sec:improved_ellipsoidal:local_evidence}

The reliable identification of isolated modes $\{M_l\}$ allows one to
evaluate the local evidence associated with each mode much more
efficiently and robustly than the methods presented in FH08.  Suppose
the $l^{\rm th}$ mode $M_l$ contains the points $\{\mathbfit{u}_j\}$
$(j=1,\cdots,n_l)$. In the simplest approach, the local evidence of this mode
is given by
\begin{equation}
\mathcal{Z}_l = \sum_{j=1}^{n_l} \mathcal{L}_jw_j,
\label{eq:localz1}
\end{equation}
where (as in Eq.~\ref{eq:finaldz}) $w_j = X_M/N$ for each active
point in $M_l$, and for each inactive points
$w_j=\tfrac{1}{2}(X_{i-1}-X_{i+1})$, in which $i$ is the nested
sampling iteration at which the inactive point was discarded. In a
similar manner, the posterior inferences resulting from the $l^{\rm th}$
mode are obtained by weighting each point in $M_l$ by
$p_j=\mathcal{L}_jw_j/Z_l$.

As outlined in FH08, however, there remain some problems with this
approach for modes that are sufficiently close to one another in the
parameter space that they are only identified as isolated regions once
the algorithm has proceeded to likelihood values somewhat larger than
the value at which the modes actually separate.  The `local' evidence
of each mode will then be underestimated by Eq.~\ref{eq:localz1}.  In
such cases, this problem can be overcome by also making use of the
points contained in the inactive groups at the end of the nested
sampling process, as follows.

For each mode $M_l$, expression Eq.~\ref{eq:localz1} for the local
evidence is replaced by
\begin{equation}
\mathcal{Z}_l = \sum_{j=1}^{n_l} \mathcal{L}_jw_j + \sum_g \mathcal{L}_gw_g\alpha^{(l)}_g,
\label{eq:localz2}
\end{equation}
where the additional summation over $g$ includes all the points in the
inactive groups, the weight $w_g=\tfrac{1}{2}(X_{i-1}-X_{i+1})$, where
$i$ is the nested sampling iteration in which the $g^{\rm th}$ point was
discarded, and the additional factors $\alpha^{(l)}_g$ are calculated
as set out below. Similarly, posterior inferences from the $l^{\rm th}$ mode
are obtained by weighting each point in $M_l$ by $p_j=\mathcal{L}_jw_j/Z_l$ and
each point in the inactive groups by $p_g=\mathcal{L}_gw_g\alpha^{(l)}_g/Z_l$.

The factors $\alpha^{(l)}_g$ can be determined in a number of
ways. The most straightforward approach is essentially to reverse the
process illustrated in Fig.~\ref{fig:groups}, as follows. Each mode
$M_l$ is simply an active group $G$ that has been renamed. Moreover,
one can identify the inactive group $G'$ that split to form $G$ at the
nested sampling iteration $i$. All points in the inactive group $G'$
are then assigned the factor
\begin{equation}
\alpha_g^{(l)} = \frac{n^{(A)}_G(i)}{n^{(A)}_{G'}(i)}, 
\label{eq:alphadef1}
\end{equation}
where $n^{(A)}_G(i)$ is the number of active points in $G$ at nested
sampling iteration $i$, and similarly for $n^{(A)}_{G'}(i)$.  Now, the
group $G'$ may itself have been formed when an inactive group
$G''$ split at an eariler nested sampling iteration $i' < i$, 
in which case all the points in $G''$ are assigned the factor
\begin{equation}
\alpha_g^{(l)} = \frac{n^{(A)}_G(i)}{n^{(A)}_{G'}(i)}
\frac{n^{(A)}_{G'}(i')}{n^{(A)}_{G''}(i')}.
\label{eq:alphadef2}
\end{equation}
The process is continued until the recursion terminates. Finally, all points
in inactive groups not already assigned have $\alpha_g^{(l)}=0$.

As a concrete example, consider $M_2=G_5$ in Fig.~\ref{fig:groups}.
In this case, the factors assigned to the members of all the inactive
groups $G_1$, $G_3$ and $G_4$ are
\begin{equation}
\alpha^{(2)}_g = 
\begin{cases}
{\displaystyle \frac{n^{(A)}_{G_5}(i_2)}{n^{(A)}_{G_3}(i_2)}_{\phantom{\int}}}
& \mbox{for $g \in G_3$} \\
{\displaystyle\frac{n^{(A)}_{G_5}(i_2)}{n^{(A)}_{G_3}(i_2)}
\frac{n^{(A)}_{G_3}(i_1)}{n^{(A)}_{G_1}(i_1)}_{\phantom{\int}}}
& \mbox{for $g \in G_1$} \\
0 & \mbox{for $g \in G_4$} 
\end{cases}
\end{equation}

It is easy to check that the general prescription (Eqs.~\ref{eq:alphadef1} and \ref{eq:alphadef2}) ensures
that
\begin{equation}
\sum_{l=1}^L \mathcal{Z}_l = \mathcal{Z},
\label{eq:checksum}
\end{equation}
i.e. the sum of the local evidences for each mode is equal to the
global evidence.  An alternative method for setting the factors
$\alpha^{(l)}_g$, for which Eq.~\ref{eq:checksum} again holds, is to use
a mixture model to analyse the full set of points (active and
inactive) produced, as outlined in Appendix A.

\section{Applications}\label{sec:applications}

In this section we apply the {\sc MultiNest} algorithm described above
to two toy problems to demonstrate that it indeed calculates the
Bayesian evidence and makes posterior inferences accurately and
efficiently. These toy examples are chosen to have features that
resemble those that can occur in real inference problems
in astro- and particle physics.

\subsection{Toy model 1: egg-box likelihood}\label{sec:applications:eggbox}

We first demonstrate the application of {\sc MultiNest} to a highly
multimodal two-dimensional problem, for which the likelihood resembles
an egg-box. The un-normalized likelihood is defined as
\begin{equation}
\mathcal{L}(\theta_1,\theta_2) = \exp \displaystyle
\left[2+\cos\left(\frac{\theta_1}{2}\right) 
\cos\left(\frac{\theta_2}{2}\right)\right]^5,
\label{eq:eggbox}
\end{equation}
and we assume a uniform prior $\mathcal{U}(0,10\pi)$ for both
$\theta_1$ and $\theta_2$.
\begin{figure*}
\psfrag{x}{$x$}
\psfrag{y}{$y$}
\psfrag{z}{$\log(\mathcal{L})$}
\begin{center}
\subfigure[]{\includegraphics[width=1\columnwidth]{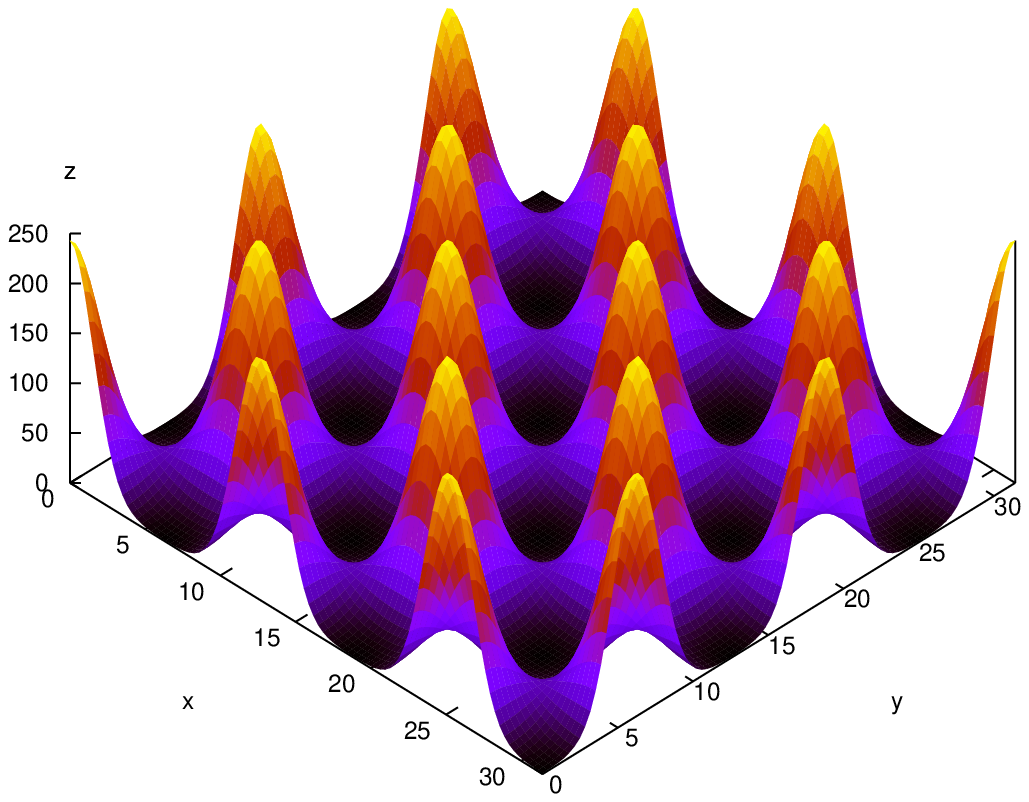}}\hspace{0.3cm}
\subfigure[]{\includegraphics[width=1\columnwidth]{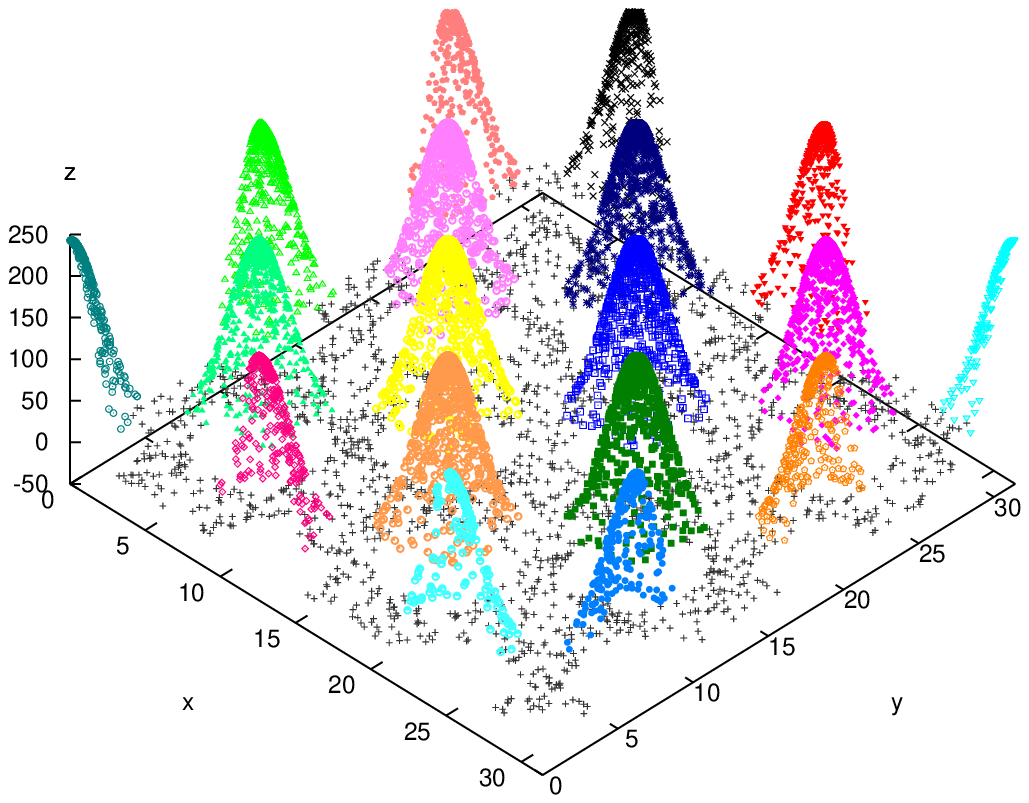}}
\caption{Toy model 1: (a) two-dimensional plot of the likelihood
  function defined in Eq.~\ref{eq:eggbox}; (b) dots denoting the
  points with the lowest likelihood at successive iterations of the
  {\sc MultiNest} algorithm. Different colours denote points assigned
  to different isolated modes as the algorithm progresses.}
\label{fig:eggbox}
\end{center}
\end{figure*}
A plot of the log-likelihood is shown in Fig.~\ref{fig:eggbox} and the
the prior ranges are chosen such that some of the modes are
truncated. Hence, although only two-dimensional, this toy example is a
particularly challenging problem, not only for mode identification but
also for evaluating the local evidence of each mode
accurately. Indeed, even obtaining posterior samples efficiently from
such a distribution can present a challenge for standard
Metropolis--Hastings MCMC samplers. We note that distributions of this
sort can occur in astronomical object detection applications (see
FH08).

Owing to the highly multimodal nature of this problem, we use 2000
active points. The results obtained with {\sc MultiNest} are
illustrated in Fig.~\ref{fig:eggbox}, in which the dots show the
points with the lowest likelihood at successive iterations of the
nested sampling process, and different colours indicate points
assigned to different isolated modes as the algorithm progresses. {\sc
  MultiNest} required $\sim 30,000$ likelihood evaluations and
evaluated the global log-evidence value to be $235.86 \pm 0.06$, which
compares favourably with the log-evidence value of $235.88$ obtained
through numerical integration on a fine grid. The local log-evidence values of each mode,
calculated through numerical integration on a fine grid (denoted as
`true $\log(\mathcal{Z})$') and using {\sc MultiNest} are listed in
Table~\ref{tab:eggbox}. We see that there is good agreement between
the two estimates.

\begin{table}
\begin{center}
\begin{tabular}{ccc}
\hline
Mode & true local $\log(\mathcal{Z})$  & {\sc MultiNest} local $\log(\mathcal{Z})$ \\
\hline
1  & $233.33$ & $233.20 \pm 0.08$ \\
2  & $233.33$ & $233.10 \pm 0.06$ \\
3  & $233.33$ & $233.48 \pm 0.05$ \\
4  & $233.33$ & $233.43 \pm 0.05$ \\
5  & $233.33$ & $233.65 \pm 0.05$ \\
6  & $233.33$ & $233.27 \pm 0.05$ \\
7  & $233.33$ & $233.14 \pm 0.06$ \\
8  & $233.33$ & $233.81 \pm 0.04$ \\
9  & $232.64$ & $232.65 \pm 0.12$ \\
10 & $232.64$ & $232.43 \pm 0.16$ \\
11 & $232.64$ & $232.11 \pm 0.14$ \\
12 & $232.64$ & $232.44 \pm 0.11$ \\
13 & $232.64$ & $232.68 \pm 0.11$ \\
14 & $232.64$ & $232.84 \pm 0.09$ \\
15 & $232.64$ & $233.02 \pm 0.09$ \\
16 & $232.64$ & $231.65 \pm 0.29$ \\
17 & $231.94$ & $231.49 \pm 0.27$\\
18 & $231.94$ & $230.46 \pm 0.36$ \\
\hline
\end{tabular}
\caption{The local log-evidence values of each mode for the toy model
  1, described in Section \ref{sec:applications:eggbox}, calculated
  through numerical integration on a fine grid (the `true
  $\log(\mathcal{Z})$') and using the {\sc MultiNest} algorithm.}
\label{tab:eggbox}
\end{center}
\end{table}

\subsection{Toy model 2: Gaussian shells likelihood}\label{sec:applications:shells}

We now illustrate the capabilities of our {\sc MultiNest} in sampling
from a posterior containing multiple modes with pronounced (curving)
degeneracies, and extend our analysis to parameters spaces of high
dimension. 

\begin{figure*}
\begin{center}
\subfigure[]{\includegraphics[width=0.8\columnwidth]{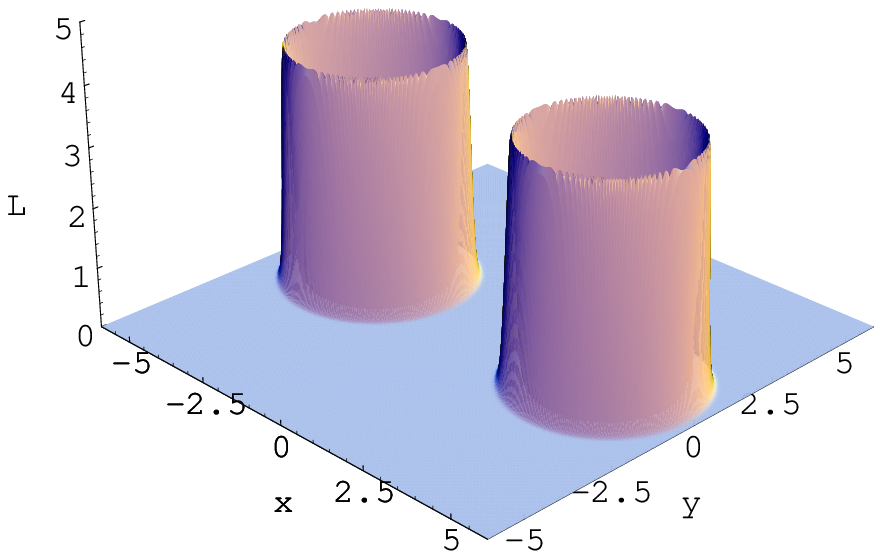}}\hspace{0.3cm}
\subfigure[]{\includegraphics[width=0.95\columnwidth]{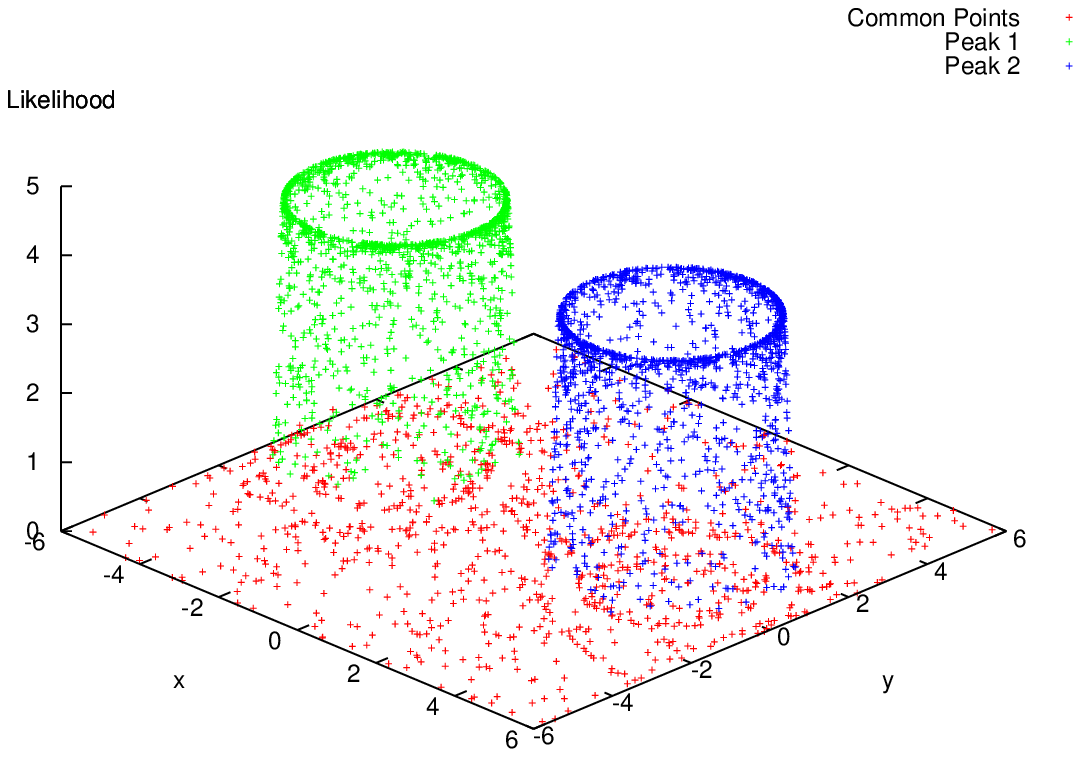}}
\caption{Toy model 2: (a) two-dimensional plot of the likelihood
  function defined in Eqs.~(\ref{eq:gshellL}) and (\ref{eq:gshell});
  (b) dots denoting the points with the lowest likelihood at
  successive iterations of the {\sc MultiNest} algorithm. Different
  colours denote points assigned to different isolated modes as the
  algorithm progresses.}
\label{fig:gshells}
\end{center}
\end{figure*}

Our toy problem here is the same one used in FH08 and \citet{bank}. The likelihood function in this model is
defined as,
\begin{equation}
\mathcal{L}(\btheta)={\rm circ}(\btheta;\bmath{c}_1,r_1,w_1)
+{\rm circ}(\btheta;\bmath{c}_2,r_2,w_2),
\label{eq:gshellL}
\end{equation}
where
\begin{equation}
{\rm circ}(\btheta;\bmath{c},r,w)=\frac{1}{\sqrt{2\pi
w^2}}\exp\left[-\frac{(\left|\btheta-\bmath{c}\right|-r)^2}{2w^2}\right].
\label{eq:gshell}
\end{equation}
In two dimensions, this toy distribution represents two well separated
rings, centred on the points $\bmath{c}_1$ and $\bmath{c}_2$
respectively, each of radius $r$ and with a Gaussian radial profile of
width $w$ (see Fig.~\ref{fig:gshells}). With a sufficiently small $w$
value, this distribution is representative of the likelihood functions
one might encounter in analysing forthcoming particle physics
experiments in the context of beyond-the-Standard-Model paradigms; in
such models the bulk of the probability lies within thin sheets or
hypersurfaces through the full parameter space.

We investigate the above distribution up to a $30$-dimensional
parameter space $\mathbf{\Theta}$ with {\sc MultiNest}. In all cases,
the centres of the two rings are separated by $7$ units in the
parameter space, and we take $w_1=w_2=0.1$ and $r_1=r_2=2$. We make
$r_1$ and $r_2$ equal, since in higher dimensions any slight
difference between these two values would result in a vast difference
between the volumes occupied by the rings and consequently the ring
with the smaller $r$ value would occupy a vanishingly small fraction
of the total probability volume, making its detection almost
impossible. It should also be noted that setting $w=0.1$ means the
rings have an extremely narrow Gaussian profile and hence they
represent an `optimally difficult' problem for our ellipsoidal nested
sampling algorithm, since many tiny ellipsoids are required to obtain
a sufficiently accurate representation of the iso-likelihood
surfaces. For the two-dimensional case, with the parameters described
above, the likelihood is shown in Fig.~\ref{fig:gshells}.

\begin{table*}
\begin{center}
\begin{tabular}{rrrrrrrrr}
\hline
&\multicolumn{2}{|c|}{Analytical}&\multicolumn{3}{|c|}{{\sc MultiNest}}\\
$D$&log$(\mathcal{Z})$&local log$(\mathcal{Z})$&log$(\mathcal{Z})$&local log$(\mathcal{Z}_1)$&local log$(\mathcal{Z}_2)$\\
\hline
$2 $ & $ -1.75$ & $ -2.44$ & $ -1.72 \pm 0.05$ & $ -2.28 \pm 0.08$ & $ -2.56 \pm 0.08$\\
$5 $ & $ -5.67$ & $ -6.36$ & $ -5.75 \pm 0.08$ & $ -6.34 \pm 0.10$ & $ -6.57 \pm 0.11$\\
$10$ & $-14.59$ & $-15.28$ & $-14.69 \pm 0.12$ & $-15.41 \pm 0.15$ & $-15.36 \pm 0.15$\\
$20$ & $-36.09$ & $-36.78$ & $-35.93 \pm 0.19$ & $-37.13 \pm 0.23$ & $-36.28 \pm 0.22$\\
$30$ & $-60.13$ & $-60.82$ & $-59.94 \pm 0.24$ & $-60.70 \pm 0.30$ & $-60.57 \pm 0.32$\\
\hline
\end{tabular}
\caption{The true and estimated global and local $\log(\mathcal{Z})$ for toy model 2, as a function of the
dimensions $D$ of the parameter space, using {\sc MultiNest}.}
\label{tab:gshells_evidence}
\end{center}
\end{table*}

In analysing this problem using the methods presented in FH08, we
showed that the sampling efficiency dropped significantly with
increasing dimensionality, with the efficiency being less than 2 per
cent in 10 dimensions, with almost $600,000$ likelihood evaluations
required to estimate the evidence to the required accuracy.  Using
1000 active points in {\sc MultiNest}, we list the evaluated and
analytical evidence values in Table \ref{tab:gshells_evidence}. The
total number of likelihood evaluations and the sampling efficiencies
are listed in Table \ref{tab:gshells_eff}. For comparison, we also
list the number of likelihood evaluations and the sampling
efficiencies with the ellipsoidal nested sampling method proposed in
FH08. One sees that {\sc MultiNest} requires an order of magnitude
fewer likelihood evaluations than the method of FH08.  In fact, the
relative computational cost of {\sc MultiNest} is even less than this
comparison suggests, since it no longer performs an eigen-analysis at
each iteration, as discussed in
Section \ref{sec:improved_ellipsoidal:dino}. Indeed, for this toy problem
discussed, the EM partitioning algorithm discussed in
Section \ref{sec:improved_ellipsoidal:dino} was on average called only
once per 1000 iterations of the {\sc MultiNest} algorithm.

\begin{table}
\begin{center}
\begin{tabular}{rrrrr}
\hline
&\multicolumn{2}{|c|}{from FH08}&\multicolumn{2}{|c|}{{\sc MultiNest}}\\
$D$ & $N_{\rm like}$ & Efficiency & $N_{\rm like}$ & Efficiency\\
\hline
$2$&$27,658$&$15.98\%$&$7,370$&$70.77\%$\\
$5$&$69,094$&$9.57\%$&$17,967$&$51.02\%$\\
$10$&$579,208$&$1.82\%$&$52,901$&$34.28\%$\\
$20$&$43,093,230$&$0.05\%$&$255,092$&$15.49\%$\\
$30$&$$&$$&$753,789$&$8.39\%$\\
\hline
\end{tabular}
\caption{The number of likelihood evaluations and sampling efficiency
 for the ellipsoidal nested sampling algorithm of FH08 and
{\sc MultiNest}, when applied to toy model 2 as a function of
  the dimension $D$ of the parameter space.}
\label{tab:gshells_eff}
\end{center}
\end{table}

\section{Cosmological parameter estimation and model 
selection}\label{sec:cosmology}

Likelihood functions resembling those used in our toy models do occur in real inference problems in astro- and
particle physics, such as object detection in astronomy (see e.g. Hobson \& McLachlan 2003; FH08) and analysis of
beyond-the-Standard-Model theories in particle physics phenomenology (see e.g. \citealt{Feroz:2008wr}).
Nonetheless, not all likelihood functions are as challenging and it is important to demonstrate that {\sc
MultiNest} is more efficient (and certainly no less so) than standard Metropolis--Hastings MCMC sampling even in
more straightforward inference problems.

An important area of inference in astrophysics is that of cosmological parameter estimation and model selection,
for which the likelihood functions are usually quite benign, often resembling a single, broad multivariate
Gaussian in the allowed parameter space. Therefore, in this section, we apply the {\sc MultiNest} algorithm to
analyse two related extensions of the standard cosmology model: non-flat spatial curvature and a varying equation
of state of dark energy.

The complete set of cosmological parameters and the ranges of the uniform priors assumed for them are given in
Table \ref{table:priors}, where the parameters have their usual meanings. With $\Omega_{\rm k} = 0$ and $w=-1$
this model then represents the `vanilla' $\Lambda$CDM cosmology. In addition, mirroring the recent analysis of
the WMAP 5-year (WMAP5) data \citep{Dunkley}, a Sunyaev-Zel'dovich amplitude is introduced, with a uniform prior
in the range $[0,2]$.  We have chosen three basic sets of data: CMB observations alone; CMB plus the Hubble Space
Telescope (HST) constraint on $H_0$ \citep{Freedman}; and CMB plus large scale structure (LSS) constraints on the
matter power spectrum derived from the luminous red galaxy (LRG) subset of the Sloan Digital Sky Survey (SDSS;
\citealt{SDSSI}; \citealt{SDSSII}) and the two degree field survey (2dF; \citealt{2dFI}). In addition, for the
dark energy analysis we include distance measures from supernovae Ia data \citep{Kowalski}. The CMB data
comprises WMAP5 observations \citep{WMAPIII} + higher resolution datasets from the Arcminute Cosmology Bolometer
Array (ACBAR; \citealt{ACBAR2008}) + the Cosmic Background Imager (CBI; \citealt{CBI}; \citealt{CBI_II};
\citealt{CBI_III}) + Balloon Observations of Millimetric Extragalactic Radiation and Geophysics ({\sc BOOMERanG};
\citealt{BOOMI}; \citealt{BOOMII}; \citealt{BOOMIII}).

Observations of the first CMB acoustic peak cannot in themselves constrain the spatial curvature $\Omega_{\rm
k}$. This could be constrained using angular scale of the first acoustic peak coupled with a knowledge of the
distance to the last scattering surface, but the latter is a function of the entire expansion history of the
universe and so there is a significant degeneracy between $\Omega_{\rm k}$ and the Hubble parameter $H(z)$. This
dependence, often termed the `geometric degeneracy', can be broken, however, since measurements at different
redshifts can constrain a sufficiently different functions of $\Omega_{\rm k}$ and $H$. Thus, the combination of
CMB data with measurements of the acoustic peak structure in the matter power spectrum derived from large scale
structure surveys such as the LRG subset of Sloan can place much tighter constraints on curvature than with
either alone (see e.g. \citealt{Eisenstein}; \citealt{TegmarkII}; \citealt{Komatsu}).

Inflation generically predicts a flat universe \citep{Guth}. The tightest current constraint suggests
$\Omega_{\rm k} \approx 10^{-2}$, whereas inflation lasting over 60 e-folds would produce flatness at the level
of $10^{-5}$ \citep{Komatsu}. Thus, at present, the data is not capable of refuting or confirming such an
inflationary picture. From a Bayesian point of view, however, one can still assess whether the data currently
prefers the inclusion of such a physical parameter. 
\begin{table}
\begin{center}
\begin{tabular}{|c||c||c|} 
\hline 
$0.018$ & $\leq \Omega_{\rm b} h^2 \leq$ &$0.032$\\
$0.04$ & $\leq \Omega_{\rm cdm} h^2 \leq$ &$0.16$\\
$0.98$ & $\leq \Theta \leq$ &$1.1$\\ 
$0.01$ & $\leq \tau \leq$ &$0.5$\\
$-0.1$ & $\leq \Omega_{\rm k} \leq$ &$0.1$\\
$-1.5$ & $\leq w \leq$ &$-0.5$\\
$0.8$  & $\leq n_{\rm s} \leq$ &$1.2$\\
$2.6$  & $\leq \log[10^{10} A_{\rm s}] \leq$ &$4.2$\\ 
\hline
\end{tabular}
\caption{Cosmological parameters and uniform priors ranges for the vanilla $\Lambda$CDM model, plus spatial
curvature $\Omega_{\rm k}$ and dark energy equation of state parameter $w$.} 
\label{table:priors}
\end{center}
\end{table}

The algorithmic parameters of {\sc MultiNest} were appropriately chosen given our a priori knowledge of the
uni-modal form of typical cosmological posteriors, the dimensionality of the problem and some empirical testing. 
The number of active points was set to $N=400$ and a sampling efficiency $e$ of 0.3 means that MPI
parallelisation across 4 CPUs is optimal (with a further 4 openmp threads per MPI CPU used by {\sc CAMB}'s
multithreading facility).  This represents a relatively modest computational investment. All of the inferences
obtained in this section required between 40,000 and 50,000 likelihood evaluations.

\subsection{Results: spatial curvature}\label{sec:cosmology:results1}

\begin{figure}
\begin{center}
\includegraphics[width=0.8\linewidth]{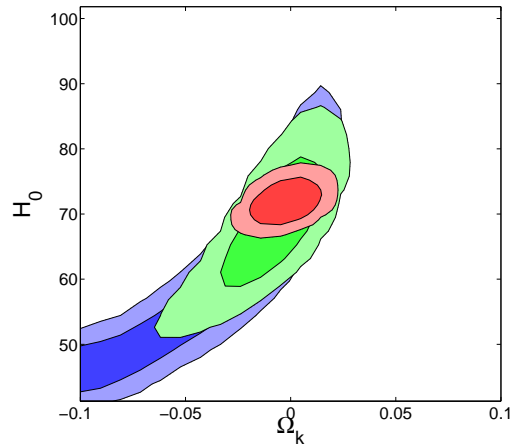}
\caption{Breaking of the `geometric' degeneracy in CMB data (blue) via
  the addition of HST (green) and large scale structure data (red).}
\label{figure:2D_curvature}
\end{center}
\end{figure}

Fig. \ref{figure:2D_curvature}. illustrates the progressively tighter constraints placed on $\Omega_{\rm k}$
and $H_0$ produced by combining CMB data with other cosmological probes of large scale structure. The geometric
degeneracy is clearly unbroken with CMB data alone, but the independent constraints on $H_0$ by HST are seen to
tighten the constraint somewhat. Including LSS data, specifically the LRG data, markedly reduces the uncertainty
on the curvature so that at 1-$\sigma$ we can limit the curvature range to $-0.043 \leq \Omega_{\rm k} \leq
0.004$. The asymmetry of this constraint leaves a larger negative tail of $\Omega_{\rm k}$ resulting in a mean
value that is only slightly closed. However, even these most stringent parameter constraints available, we see no
statistically significant deviation from a spatially flat universe. The Bayesian evidence, in penalising
excessive model complexity should tell us whether relaxing the constraint on flatness is preferred by the data.
Our results (Table \ref{table:evidence}) very clearly rule out the necessity for such an addition in anything
other than with CMB data alone. This implies that the inclusion of spatial curvature is an unnecessary
complication in cosmological model building, given the currently available data.
\begin{table}
\begin{center}
\begin{tabular}{|c||c||c|}
    \hline
 \textbf{Dataset} $\backslash$ \textbf{Model}  & vanilla + $\Omega_{\rm k}$ & vanilla + $\Omega_{\rm k}$ + $w$\\
    \hline
 CMB alone	&  $-0.29 \pm 0.27$ & -\\
 CMB + HST  &  $-1.56 \pm 0.27$ & -\\
 ALL & $-2.92 \pm 0.27$ & $-1.29 \pm 0.27$\\
    \hline
\end{tabular}
\caption{Differences of $\log$ evidences for both models and the three
  datasets described in the text.[Negative (Positive) values represent
    lower (higher) preference for the parameterisation change]}
\label{table:evidence}
\end{center}
\end{table}
%

\subsection{Results: varying equation of state of dark energy}\label{sec:cosmology:results2}

The properties of the largest universal density component are still largely unknown, yet such a component seems
crucial for the universe to be spatially flat.  It has thus been argued by some (\citealt{Wright} \&
\citealt{TegmarkII}) that it is inappropriate to assume spatial flatness when attempting to vary the properties
of the dark energy component beyond those of a simple cosmological constant. Here we allow for the variation of
the dark energy equation of state parameter $w$. We will therefore proceed by placing joint constraints on both
$w$ and $\Omega_{\rm k}$, as performed in \citet{Komatsu}. Once again we encounter a serious degeneracy, this
time between $\Omega_{\rm k}$ and $w$. With the geometric degeneracy, combining cosmological observations of the
universe at different redshifts was sufficient to break the dependence, but when dark energy is dynamical we
\emph{must} use at least a third, independent data set. In this analysis, we have therefore included distance
measures from type Ia supernovae observations \citep{Kowalski}.  Using this combination of data produces
impressively tight constraints on both $w$ and $\Omega_{\rm k}$; indeed the resulting constraints on the spatial
curvature are tighter than those obtained in the previous section, for which $w$ was held constant at $w=-1$. 
This is primarily due to the near orthogonality of the constraints provided by supernovae and the CMB.  Once
again we find little evidence to support a departure from the basic vanilla cosmology (see Table
\ref{table:evidence}). To within estimated uncertainty, the Bayesian evidence is at least one log unit greater
for a flat universe with dark energy in the form of a cosmological constant.
\begin{figure}
\begin{center}
\includegraphics[width=0.8\linewidth]{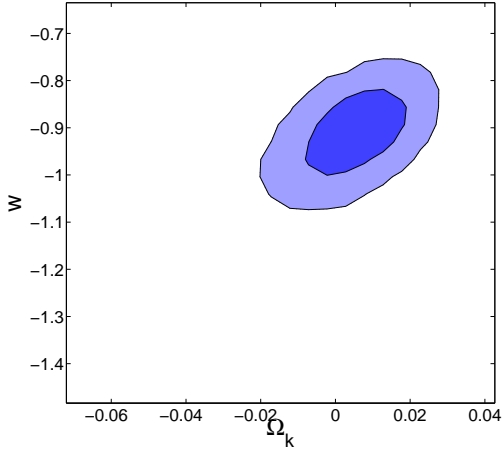}
\caption{Joint constraints on universal geometry $\Omega_{\rm k}$ and the equation of state of dark energy $w$
using WMAP5 + HST + LSS + supernovae data.}
\label{figure:2D_dark_energy}
\end{center}
\end{figure}
%

\subsection{Comparison of {\sc MultiNest} and MCMC `quick look' parameter constraints}\label{sec:MCMC_comparison}

The above results are in good agreement with those found by
\citet{Komatsu} using more traditional MCMC methods, and indicate that
      {\sc MultiNest} has produced reliable inferences, both in terms
      of the estimated evidence values and the derived posterior
      parameter constraints.  It is often the case, however, that
      cosmologists wish only to obtain a broad idea of the posterior
      distribution of parameters, using short MCMC chains and hence
      relatively few likelihood evaluations.  In this section, we show
      that the {\sc MultiNest} algorithm can also perform this task by
      setting the number of active points, $N$, to a smaller value.

In order to illustrate this functionality, we analyse the WMAP5 CMB
data-set in the context of the vanilla $\Lambda$CDM cosmology using
both {\sc MultiNest} and the publicly available CosmoMC
\citet{cosmomc} package, which uses an MCMC sampler based on a
tailored version of the Metropolis--Hastings method. We imposed
uniform priors on all the parameters. The prior ranges for
$\Omega_{\rm b} h^2, \Omega_{\rm cdm} h^2, \Theta$, and $\tau$ is
listed in Table~\ref{table:priors}. In addition, $n_{\rm s}$ was
allowed to vary between 0.8 and 1.2, $\log(10^{10} A_{\rm s})$ between
2.6 and 4.2 and the Sunyaev--Zel'dovich amplitude between 0 and 2.

To facilitate later comparisons, we first obtained an accurate
determination of the posterior distribution using the traditional
method by running CosmoMC to produce 4 long MCMC chains. The
Gelman--Rubin statistic $R$ returned by CosmoMC indicated that the
chains had converged to within $R \approx 1.1$ after about 6000 steps
per chain, resulting in $\sim$ 24,000 likelihood evaluations. To be
certain of determining the `true' posterior to high accuracy, we then
ran the MCMC chains for a further 6,000 samples per chain, resulting
in a total of 48,000 likelihood evaluations, at which point the
convergence statistic was $R \approx 1.05$.

\begin{figure}
\psfrag{xl1}{$\Omega_{\rm b} h^2$}
\psfrag{xl2}{$\Omega_{\rm cdm} h^2$}
\psfrag{xl3}{$\Theta$}
\psfrag{xl4}{$\tau$}
\psfrag{xl5}{$n_{\rm s}$}
\psfrag{xl6}{$\log(10^{10} A_{\rm s})$}
\begin{center}
\includegraphics[width=1\columnwidth]{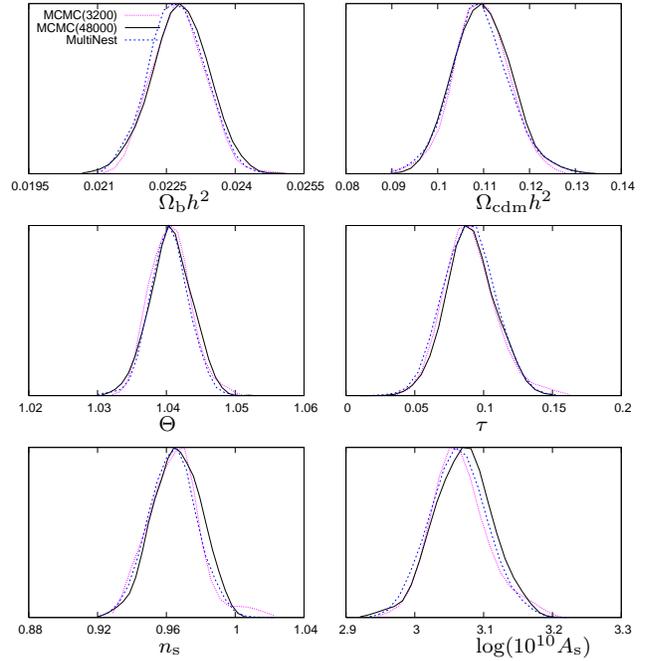}
\caption{1-D marginalized posteriors for the flat-$\Lambda$CDM cosmology
obtained with: CosmoMC using 48,000 likelihood evaluations (solid black);
  CosmoMC using 3,200 likelihood evaluations (dotted pink); and {\sc
    MultiNest} using 3,100 likelihood evaluations (dashed blue).}
\label{fig:mcmc_compar}
\end{center}
\end{figure}

As stated above, however, one often wishes to obtain only a
`quick-and-dirty' estimate of the posterior in the first stages of the
data analysis. As might be typical of such analyses, we ran CosmoMC
again using 4 MCMC chains, but with only 800 steps per chain,
resulting in a total of 3,200 likelihood evaluations.  As a
comparison, we also ran the {\sc MultiNest} algorithm using only 50
active points and with the sampling efficiency $e$ set to unity;
this required a total of 3,100 likelihood evaluations. In
Fig.~\ref{fig:mcmc_compar}, we plot the 1-D marginalized posteriors for
derived from the two analyses, together with the results of the longer
CosmoMC analysis described above.  It is clear from these plots that
both the {\sc MultiNest} and MCMC `quick-look' results compare well
with the `true' posterior obtained from the more costly rigorous
analysis.

\section{Discussion and conclusions}\label{sec:conclusions}

We have described a highly efficient Bayesian inference tool, called
{\sc MultiNest}, which we have now made freely available for academic
purposes as a plugin that is easily incorporated into the {\sc
  CosmoMC} software.  On challenging toy models that resemble real
inference problems in astro- and particle physics, we have
demonstrated that {\sc MultiNest} produces reliable estimates of the
evidence, and its uncertainty, and accurate posterior inferences from
distributions with multiple modes and pronounced curving degeneracies
in high dimensions. We have also demonstrated in cosmological
inference problems that {\sc MultiNest} produces accurate parameter
constraints on similar time scales to standard MCMC methods
\emph{and}, with negligible extra computational effort, also yields
very accurate Bayesian evidences for model selection. As a
cosmological application we have considered two extensions of the
basic vanilla $\Lambda$CDM cosmology: non-zero spatial curvature and a
varying equation of state of dark energy. Both extensions are
determined to be unnecessary for the modelling of existing data via
the evidence criterion, confirming that with the advent of five years
of WMAP observations the data is still satisfied by a $\Lambda$CDM
cosmology.

As a guide for potential users, we conclude by noting that the {\sc
  MultiNest} algorithm is controlled by two main parameters: (i) the
number of active points $N$; and (ii) the maximum efficiency $e$ (see
Section \ref{sec:improved_ellipsoidal:dino}). These values can be chosen
quite easily as outlined below. First, $N$ should be large enough
that, in the initial sampling from the full prior space, there is a
high probability that at least one point lies in the `basin of
attraction' of each mode of the posterior. In later iterations, active
points will then tend to populate these modes. It should be
remembered, of course, that $N$ must always exceed the dimensionality
$D$ of the parameter space. Also, in order to calculate the evidence
accurately, $N$ should be sufficiently large so that all the regions
of the parameter space are sampled adequately. For parameter
estimation only, one can use far fewer active points. For cosmological
data analysis, we found $400$ and $50$ active points to be adequate for
evidence evaluation and parameter estimation respectively.
The parameter $e$ controls the sampling volume
at each iteration, which is equal to the sum of the
volumes of the ellipsoids enclosing the active point set. For parameter
estimation problems, $e$ should be set to $1$ to obtain maximum
efficiency without undersampling or to a lower value if one wants to
get a general idea of the posterior very quickly. For evidence
evaluation in cosmology, we found setting $e \sim 0.3$ ensures an
accurate evidence value.

\section*{Acknowledgements}

This work was carried out largely on the Cambridge High Performance Computing Cluster Darwin and the
authors would like to thank Dr. Stuart Rankin for computational assistance. FF is supported by studentships from the Cambridge
Commonwealth Trust, Isaac Newton and the Pakistan Higher Education Commission Fellowships. MB is supported by
STFC.

\bibliographystyle{mn2e}
\bibliography{multinest2}

\begin{thebibliography}{}

\bibitem[\protect\citeauthoryear{Allanach \& Lester}{Allanach \&
  Lester}{2007}]{bank}
Allanach B.~C.,  Lester C.~G., , 2007, Sampling using a `bank' of clues

\bibitem[\protect\citeauthoryear{Bassett, Corasaniti \& Kunz}{Bassett
  et~al.}{2004}]{Bassett04}
Bassett B.~A.,  Corasaniti P.~S.,    Kunz M.,  2004, Astrophys. J., 617, L1

\bibitem[\protect\citeauthoryear{{Beltr{\'a}n}, {Garc{\'{\i}}a-Bellido},
  {Lesgourgues}, {Liddle} \& {Slosar}}{{Beltr{\'a}n} et~al.}{2005}]{Beltran05}
{Beltr{\'a}n} M.,  {Garc{\'{\i}}a-Bellido} J.,  {Lesgourgues} J.,  {Liddle}
  A.~R.,    {Slosar} A.,  2005, \prd, 71, 063532

\bibitem[\protect\citeauthoryear{{Bridges}, {Lasenby} \& {Hobson}}{{Bridges}
  et~al.}{2006}]{Bridges06a}
{Bridges} M.,  {Lasenby} A.~N.,    {Hobson} M.~P.,  2006, \mnras, 369, 1123

\bibitem[\protect\citeauthoryear{{CBI Supplementary Data}}{{CBI Supplementary
  Data}}{2006}]{CBI_III}
{CBI Supplementary Data} 2006

\bibitem[\protect\citeauthoryear{Cole et~al.,}{Cole  et~al.}{2005}]{2dFI}
Cole S.,  et~al., 2005, Mon. Not. Roy. Astron. Soc., 362, 505

\bibitem[\protect\citeauthoryear{Dempster, Laird \& Rubin}{Dempster
  et~al.}{1977}]{Dempster77}
Dempster A.~P.,  Laird N.~M.,    Rubin D.~B.,  1977, Journal of the Royal
  Statistical Society. Series B (Methodological), 39, 1

\bibitem[\protect\citeauthoryear{{Dunkley}, {Komatsu}, {Nolta}, {Spergel},
  {Larson}, {Hinshaw}, {Page}, {Bennett}, {Gold}, {Jarosik}, {Weiland},
  {Halpern}, {Hill}, {Kogut}, {Limon}, {Meyer}, {Tucker}, {Wollack} \&
  {Wright}}{{Dunkley} et~al.}{2008}]{Dunkley}
{Dunkley} J.,  {Komatsu} E.,  {Nolta} M.~R.,  {Spergel} D.~N.,  {Larson} D.,
  {Hinshaw} G.,  {Page} L.,  {Bennett} C.~L.,  {Gold} B.,  {Jarosik} N.,
  {Weiland} J.~L.,  {Halpern} M.,  {Hill} R.~S.,  {Kogut} A.,  {Limon} M.,
  {Meyer} S.~S.,  {Tucker} G.~S.,  {Wollack} E.,    {Wright} E.~L.,  2008,
  ArXiv e-prints, 803

\bibitem[\protect\citeauthoryear{{Eisenstein}, {Zehavi}, {Hogg}, {Scoccimarro},
  {Blanton}, {Nichol}, {Scranton}, {Seo}, {Tegmark}, {Zheng} \&
  {Anderson}}{{Eisenstein} et~al.}{2005}]{Eisenstein}
{Eisenstein} D.~J.,  {Zehavi} I.,  {Hogg} D.~W.,  {Scoccimarro} R.,  {Blanton}
  M.~R.,  {Nichol} R.~C.,  {Scranton} R.,  {Seo} H.-J.,  {Tegmark} M.,  {Zheng}
  Z.,    {Anderson} S.~F.,  2005, \apj, 633, 560

\bibitem[\protect\citeauthoryear{{Feroz}, {Allanach}, {Hobson}, {AbdusSalam},
  {Trotta} \& {Weber}}{{Feroz} et~al.}{2008}]{Feroz:2008wr}
{Feroz} F.,  {Allanach} B.~C.,  {Hobson} M.,  {AbdusSalam} S.~S.,  {Trotta} R.,
     {Weber} A.~M., , 2008, {Bayesian Selection of sign(mu) within mSUGRA in
  Global Fits Including WMAP5 Results}

\bibitem[\protect\citeauthoryear{{Feroz} \& {Hobson}}{{Feroz} \&
  {Hobson}}{2008}]{feroz08}
{Feroz} F.,  {Hobson} M.~P.,  2008, \mnras, 384, 449

\bibitem[\protect\citeauthoryear{{Freedman}, {Madore}, {Gibson}, {Ferrarese},
  {Kelson}, {Sakai}, {Mould}, {Kennicutt} Jr., {Ford}, {Graham}, {Huchra},
  {Hughes}, {Illingworth}, {Macri} \& {Stetson}}{{Freedman}
  et~al.}{2001}]{Freedman}
{Freedman} W.~L.,  {Madore} B.~F.,  {Gibson} B.~K.,  {Ferrarese} L.,  {Kelson}
  D.~D.,  {Sakai} S.,  {Mould} J.~R.,  {Kennicutt} Jr. R.~C.,  {Ford} H.~C.,
  {Graham} J.~A.,  {Huchra} J.~P.,  {Hughes} S.~M.~G.,  {Illingworth} G.~D.,
  {Macri} L.~M.,    {Stetson} P.~B.,  2001, \apj, 553, 47

\bibitem[\protect\citeauthoryear{{Guth}}{{Guth}}{1981}]{Guth}
{Guth} A.~H.,  1981, \prd, 23, 347

\bibitem[\protect\citeauthoryear{{Hinshaw}, {Weiland}, {Hill}, {Odegard},
  {Larson}, {Bennett}, {Dunkley} \& {Gold}}{{Hinshaw} et~al.}{2008}]{WMAPIII}
{Hinshaw} G.,  {Weiland} J.~L.,  {Hill} R.~S.,  {Odegard} N.,  {Larson} D.,
  {Bennett} C.~L.,  {Dunkley} J.,    {Gold} B.,  2008, ArXiv e-prints, 803

\bibitem[\protect\citeauthoryear{{Hobson}, {Bridle} \& {Lahav}}{{Hobson}
  et~al.}{2002}]{Hobson02}
{Hobson} M.~P.,  {Bridle} S.~L.,    {Lahav} O.,  2002, \mnras, 335, 377

\bibitem[\protect\citeauthoryear{{Hobson} \& {McLachlan}}{{Hobson} \&
  {McLachlan}}{2003}]{Hobson03}
{Hobson} M.~P.,  {McLachlan} C.,  2003, \mnras, 338, 765

\bibitem[\protect\citeauthoryear{Jones et~al.,}{Jones  et~al.}{2006}]{BOOMII}
Jones W.~C.,  et~al., 2006, \apj, 647, 823

\bibitem[\protect\citeauthoryear{{Komatsu}, {Dunkley}, {Nolta}, {Bennett},
  {Gold}, {Hinshaw}, {Jarosik}, {Larson}, {Limon}, {Page}, {Spergel},
  {Halpern}, {Hill}, {Kogut}, {Meyer}, {Tucker}, {Weiland}, {Wollack} \&
  {Wright}}{{Komatsu} et~al.}{2008}]{Komatsu}
{Komatsu} E.,  {Dunkley} J.,  {Nolta} M.~R.,  {Bennett} C.~L.,  {Gold} B.,
  {Hinshaw} G.,  {Jarosik} N.,  {Larson} D.,  {Limon} M.,  {Page} L.,
  {Spergel} D.~N.,  {Halpern} M.,  {Hill} R.~S.,  {Kogut} A.,  {Meyer} S.~S.,
  {Tucker} G.~S.,  {Weiland} J.~L.,  {Wollack} E.,    {Wright} E.~L.,  2008,
  ArXiv e-prints, 803

\bibitem[\protect\citeauthoryear{Kowalski et~al.,}{Kowalski
  et~al.}{2008}]{Kowalski}
Kowalski M.,  et~al., 2008

\bibitem[\protect\citeauthoryear{{Lewis} \& {Bridle}}{{Lewis} \&
  {Bridle}}{2002}]{cosmomc}
{Lewis} A.,  {Bridle} S.,  2002, \prd, 66, 103511

\bibitem[\protect\citeauthoryear{{Liddle}}{{Liddle}}{2007}]{Liddle07}
{Liddle} A.~R.,  2007, ArXiv Astrophysics e-prints

\bibitem[\protect\citeauthoryear{Lu, Choi, Wang \& Kim}{Lu et~al.}{2007}]{Lu07}
Lu L.,  Choi Y.-K.,  Wang W.,    Kim M.-S.,  2007, Computer Graphics Forum, 26,
  329

\bibitem[\protect\citeauthoryear{{Mackay}}{{Mackay}}{2003}]{MacKay}
{Mackay} D.~J.~C.,  2003, {Information Theory, Inference and Learning
  Algorithms}.
Information Theory, Inference and Learning Algorithms, by David J.~C.~MacKay,
  pp.~640.~ISBN 0521642981.~Cambridge, UK: Cambridge University Press, October
  2003.

\bibitem[\protect\citeauthoryear{{Marshall}, {Hobson} \& {Slosar}}{{Marshall}
  et~al.}{2003}]{Marshall03}
{Marshall} P.~J.,  {Hobson} M.~P.,    {Slosar} A.,  2003, \mnras, 346, 489

\bibitem[\protect\citeauthoryear{Montroy et~al.,}{Montroy
  et~al.}{2006}]{BOOMIII}
Montroy T.~E.,  et~al., 2006, \apj, 647, 813

\bibitem[\protect\citeauthoryear{{Mukherjee}, {Parkinson} \&
  {Liddle}}{{Mukherjee} et~al.}{2006}]{Mukherjee06}
{Mukherjee} P.,  {Parkinson} D.,    {Liddle} A.~R.,  2006, \apjl, 638, L51

\bibitem[\protect\citeauthoryear{{Niarchou}, {Jaffe} \& {Pogosian}}{{Niarchou}
  et~al.}{2004}]{Niarchou04}
{Niarchou} A.,  {Jaffe} A.~H.,    {Pogosian} L.,  2004, \prd, 69, 063515

\bibitem[\protect\citeauthoryear{{\'O}~Ruanaidh \& Fitzgerald}{{\'O}~Ruanaidh
  \& Fitzgerald}{1996}]{Ruanaidh}
{\'O}~Ruanaidh J.,  Fitzgerald W.,  1996, Numerical Bayesian Methods Applied to
  Signal Processing.
Springer Verlag:New York

\bibitem[\protect\citeauthoryear{Piacentini et~al.,}{Piacentini
  et~al.}{2006}]{BOOMI}
Piacentini F.,  et~al., 2006, \apj, 647, 833

\bibitem[\protect\citeauthoryear{{Readhead}, {Mason}, {Contaldi}, {Pearson},
  {Bond}, {Myers}, {Padin}, {Sievers} \& {Cartwright}}{{Readhead}
  et~al.}{2004}]{CBI}
{Readhead} A.~C.~S.,  {Mason} B.~S.,  {Contaldi} C.~R.,  {Pearson} T.~J.,
  {Bond} J.~R.,  {Myers} S.~T.,  {Padin} S.,  {Sievers} J.~L.,    {Cartwright}
  J.~K.,  2004, \apj, 609, 498

\bibitem[\protect\citeauthoryear{Reichardt et~al.,}{Reichardt
  et~al.}{2008}]{ACBAR2008}
Reichardt C.~L.,  et~al., 2008

\bibitem[\protect\citeauthoryear{{Shaw}, {Bridges} \& {Hobson}}{{Shaw}
  et~al.}{2007}]{Shaw07}
{Shaw} J.~R.,  {Bridges} M.,    {Hobson} M.~P.,  2007, \mnras, 378, 1365

\bibitem[\protect\citeauthoryear{{Sievers J.~L. et al.}}{{Sievers J.~L. et
  al.}}{2007}]{CBI_II}
{Sievers J.~L. et al.} 2007, Astrophys. J., 660, 976

\bibitem[\protect\citeauthoryear{{Skilling}}{{Skilling}}{2004}]{Skilling04}
{Skilling} J.,  2004, in {Fischer} R.,  {Preuss} R.,   {Toussaint} U.~V.,  eds,
  American Institute of Physics Conference Series {Nested Sampling}.
pp 395--405

\bibitem[\protect\citeauthoryear{{Slosar A. et al.}}{{Slosar A. et
  al.}}{2003}]{Slosar03}
{Slosar A. et al.} 2003, \mnras, 341, L29

\bibitem[\protect\citeauthoryear{{Tegmark}, {Eisenstein}, {Strauss},
  {Weinberg}, {Blanton}, {Frieman}, {Fukugita}, {Gunn}, {Hamilton}, {Knapp},
  {Nichol} \& {Ostriker}}{{Tegmark} et~al.}{2006}]{TegmarkII}
{Tegmark} M.,  {Eisenstein} D.,  {Strauss} M.,  {Weinberg} D.,  {Blanton} M.,
  {Frieman} J.,  {Fukugita} M.,  {Gunn} J.,  {Hamilton} A.,  {Knapp} G.,
  {Nichol} R.,    {Ostriker} J.,  2006, ArXiv Astrophysics e-prints

\bibitem[\protect\citeauthoryear{Tegmark et~al.,}{Tegmark
  et~al.}{2004}]{SDSSII}
Tegmark M.,  et~al., 2004, Astrophys. J., 606, 702

\bibitem[\protect\citeauthoryear{Tegmark et~al.,}{Tegmark
  et~al.}{2006}]{SDSSI}
Tegmark M.,  et~al., 2006, Phys. Rev., D74, 123507

\bibitem[\protect\citeauthoryear{{Trotta}}{{Trotta}}{2007}]{Trotta05}
{Trotta} R.,  2007, \mnras, 378, 72

\bibitem[\protect\citeauthoryear{{Wright}}{{Wright}}{2006}]{Wright}
{Wright} E.~L.,  2006, ArXiv Astrophysics e-prints

\end{thebibliography}

\appendix

\section{Local evidence evaluation using a Gaussian mixture model}

As mentioned in Section \ref{sec:improved_ellipsoidal:local_evidence},
an alternative method for determining the local evidence and posterior
constraints associated with each identified mode $M_l$ is to analyse
the full set of samples using a mixture model.  In what follows, we
assume that the modes of the `implied' posterior
$\mathcal{P}(\mathbfit{u})$ in the unit hypercube space are each well
described by a multivariate Gaussian, leading to a Gaussian mixture
model, but the model for the mode shape can be easily changed to
another distribution. 

Let us assume that, at the end of the nested sampling process, the
full set of ${\cal N}$ (inactive and active) points obtained is
$\{\mathbfit{u}_1,\mathbfit{u}_{2},\cdots,\mathbfit{u}_{\cal N}\}$ and
$L$ modes $M_1, M_2, \cdots, M_L$ modes have been identified.  The
basic goal of the method is to assign an extra factor
$\alpha^{(l)}_j$ to {\em every} point ($j=1$ to $\mathcal{N}$), so
that the estimate (Eq.~\ref{eq:localz2}) for the local evidence associated
with the mode $M_l$ is replaced by
\begin{equation}
\mathcal{Z}_l = \sum_{j=1}^{\mathcal{N}} \mathcal{L}_jw_j\alpha^{(l)}_j,
\label{eq:localz3}
\end{equation}
where the weights $w_j$ for the inactive and active points are the
same as those used in Eq.~\ref{eq:localz2}. Similarly, posterior
inferences from the mode $M_l$ are obtained by weighting {\em every}
point ($j=1$ to $\mathcal{N}$) by $p_j=\mathcal{L}_jw_j\alpha^{(l)}_j/Z_l$.

The factors $\alpha^{(l)}_j$ are determined by modelling each mode
$M_l$ as a multivariate Gaussian with `normalised amplitude' $A_l$,
mean $\bmu_l$ and covariance matrix $\mathbf{C}_l$, such that
\begin{equation}
\mathcal{P}(\mathbfit{u}) \propto \sum_{l=1}^L A_l
G(\mathbfit{u};\bmu_l,\mathbf{C}_l),
\label{eq:gmmdef0}
\end{equation}
where the Gaussian unit-volume
$G(\mathbfit{u};\bmu_l,\mathbf{C}_l)$ is given by
\begin{equation}
G(\mathbfit{u};\bmu_l,\mathbf{\Theta}_l)=
\frac{1}{(2\pi)^{\frac{{\rm D}}{2}} |\mathbf{C}_l|^{\frac{1}{2}}}
\exp\displaystyle\left[-{\textstyle\frac{1}{2}}(\mathbfit{u}-\bmu_l)^{\rm
    T} \mathbf{C}_l^{-1} (\mathbfit{u}-\bmu_l)\right],
\label{eq:gmmdef}
\end{equation}
and the values of the parameters
$\mathbf{\Theta}\equiv(\{A_l\},\{\bmu_l\},\{\mathbf{C}_l\})$ are to be
determined from the sample points $\{\mathbfit{u}_j\}$. Since the
scaling in Eq.~\ref{eq:gmmdef0} is arbitrary, it is convenient to set
$\sum_{l=1}^L A_l = 1$.

For a given set of parameter values $\mathbf{\Theta}$, our required
factors are
\begin{eqnarray}
\alpha^{(l)}_j (\mathbf{\Theta}) & = &
\Pr(M_l|\mathbfit{u}_j,\mathbf{\Theta}) \nonumber \\
& = &  \frac{\Pr(\mathbfit{u}_j|M_l,\mathbf{\Theta})\Pr(M_l|\mathbf{\Theta})}
{\sum_{l=1}^L
  \Pr(\mathbfit{u}_j|M_l,\mathbf{\Theta})\Pr(M_l|\mathbf{\Theta})}
\nonumber \\
& = & \frac{A_lG(\mathbfit{u};\bmu_l,\mathbf{C}_l)}
{\sum_{l=1}^L A_lG(\mathbfit{u};\bmu_l,\mathbf{C}_l)}.
\label{eq:alphafactdef}
\end{eqnarray}

Our approach is to optimise the parameters $\mathbf{\Theta}$, and
hence determine the factors $\alpha^{(l)}_j (\mathbf{\Theta})$, using
an expectation-maximization (EM) algorithm. The 
algorithm is initialized by setting 
\begin{equation}
\alpha^{(l)}_j =
\begin{cases}
1 & \mbox{if $\mathbfit{u}_{j} \in M_{l}$}\\
0 & \mbox{otherwise}
\end{cases}
\label{eq:post_clstr_initial}
\end{equation}
and calculating the initial values of each $Z_l$ using
Eq.~\ref{eq:localz3}. In the M-step of the algorithm one then obtains the
maximum-likelihood estimates $\hat{\mathbf{\Theta}}$ of the
parameters.  These are easily derived (see e.g. ~\citet{Dempster77})
to be
\begin{eqnarray}
\widehat{A}_l & = & \frac{n_l}{n}\\ 
\widehat{\bmu}_l & = &
\frac{1}{n_l}\sum_{j=1}^{\mathcal{N}} \alpha^{(l)}_j\tilde{\mathbfit{u}}_j \\ 
\widehat{\mathbf{C}}_l & = & \frac{1}{n_l}\sum_{j=1}^{\mathcal{N}}
\alpha^{(l)}_j(\tilde{\mathbfit{u}}_j-\widehat{\bmu}_l)
(\tilde{\mathbfit{u}}_j-\widehat{\bmu}_l)^{\rm T},
\end{eqnarray}
where $n_l=\sum_{j=1}^{\mathcal{N}} \alpha^{(l)}_j$, $n=\sum_{l=1}^L
n_l$ and $\tilde{\mathbfit{u}}_j=\mathbfit{u}_jL_jw_j/Z_l$ are the
locally posterior-weighted sample points. In the subsequent E-step of
the algorithm on then updates the $\alpha^{(l)}_j$ values using
Eq.~\ref{eq:alphafactdef} and updates $Z_l$ using Eq.~\ref{eq:localz3}. We
further impose the constraint that $\alpha^{(l)}_j=0$ if
$\mathbfit{u}_j \notin M_l$ and its likelihood $\mathcal{L}_j$ is
greater than the lowest likelihood of the points in $M_l$. The EM
algorithm is then iterated to convergence.

\label{lastpage}

\end{document}